\documentclass{JINST}

\title{Operation and calibration of the Silicon Drift Detectors of the ALICE
experiment during the 2008 cosmic ray data taking period}

\author{
B.\ Alessandro$^a$,
S.\ Antinori$^b$,
R.\ Bala$^{c,a}$,
G.\ Batigne$^d$,
S.\ Beol\`e$^{c,a}$,
E.\ Biolcati$^{c,a}$,
N.\ Bock Garcia$^e$,
E.\ Bruna$^{c,1}$,
P.\ Cerello$^a$,
S.\ Coli$^a$,
Y.\ Corrales Morales$^{a,2}$,
F.\ Costa$^f$,
E.\ Crescio$^c$,
P.\ De Remigis$^a$,
S.\ Di Liberto$^g$,
D.\ Falchieri$^b$,
G.\ Feofilov$^h$,
W.\ Ferrarese$^c$,
E.\ Gandolfi$^{i,b}$,
C.\ Garcia$^{a,2}$,
L.\ Gaudichet$^a$,
G.\ Giraudo$^a$,
P.\ Giubellino$^a$,
T.\ J.\ Humanic$^e$,
S.\ Igolkin$^f$,
M.\ Idzik$^{c,a,3}$,
S.\ K.\ Kiprich$^j$,
A.\ Kisiel$^f$,
A.\ Kolozhvari$^h$,
I.\ Kotov$^e$,
J.\ Kral$^k$,
S.\ Kushpil$^l$,
V.\ Kushpil$^l$,
R.\ Lea$^{c,a}$,
M.\ A.\ Lisa$^e$,
M.\ I.\ Martinez$^{a,4}$,
A.\ Marzari Chiesa$^{c,a}$,
M.\ Masera$^{c,a}$,
M.\ Masetti$^{i,b}$,
G.\ Mazza$^a$,
M.\ A.\ Mazzoni$^g$,
F.\ Meddi$^{m,g}$,
L.\ M.\ Montano Zetina$^{c,a,5}$,
M.\ Monteno$^a$,
B.\ S.\ Nilsen$^{e,6}$,
D.\ Nouais$^a$,
F.\ Padilla Cabal$^{a,2}$,
V.\ Petr\'a\v cek$^k$,
M.\ G.\ Poghosyan$^{c,a}$,
F.\ Prino$^a$,
L.\ Ramello$^n$,
A.\ Rashevsky$^o$,
L.\ Riccati$^a$,
A.\ Rivetti$^a$,
S.\ Senyukov$^n$,
M.\ Siciliano$^{c,a}$,
M.\ Sitta$^n$\thanks{Corresponding author.},
M.\ A.\ Subieta Vasquez$^{c,a}$,
M.\ Sumbera$^l$,
L.\ Toscano$^a$,
F.\ Tosello$^a$
D.\ Truesdale$^e$,
G.\ M.\ Urciuoli$^g$,
A.\ Vacchi$^o$,
S.\ Vallero$^p$,
A.\ Werbrouck$^{q,a}$,
G.\ Zampa$^o$,
and G.\ Zinovjev$^r$\\
\llap{$^a$}Istituto Nazionale di Fisica Nucleare, Sezione di Torino,
           Turin, Italy\\
\llap{$^b$}Istituto Nazionale di Fisica Nucleare, Sezione di Bologna,
           Bologna, Italy\\
\llap{$^c$}Dipartimento di Fisica Sperimentale, Universit\`a di Torino,
           Turin, Italy\\
\llap{$^d$}SUBATECH, Ecole des Mines de Nantes, Universit\'e de Nantes,
           CNRS/IN2P3, Nantes, France\\
\llap{$^e$}Department of Physics, Ohio State University, Columbus, OH, USA\\
\llap{$^f$}European Organization for Nuclear Research, CERN, Geneva,
           Switzerland\\
\llap{$^g$}Istituto Nazionale di Fisica Nucleare, Sezione di Roma,
           Rome, Italy\\
\llap{$^h$}V. Fock Institute for Physics, St. Petersburg State University,
           St. Petersburg, Russia\\
\llap{$^i$}Dipartimento di Fisica, Universit\`a di Bologna, Bologna, Italy\\
\llap{$^j$}National Science Center, Kharkiv Institute of Physics and
           Technology, Kharkiv, Ukraine\\
\llap{$^k$}Faculty of Nuclear Sciences and Physical Engineering, Czech
           Technical University in Prague, Czech Republic\\
\llap{$^l$}Nuclear Physics Institute, Academy of Sciences of the Czech
           Republic, Czech Republic\\
\llap{$^m$}Dipartimento di Fisica, Universit\`a di Roma ``La Sapienza'',
           Rome, Italy\\
\llap{$^n$}Facolt\`a di Scienze, Universit\`a del Piemonte Orientale and INFN,
           Alessandria, Italy\\
\llap{$^o$}Istituto Nazionale di Fisica Nucleare, Sezione di Trieste,
           Trieste, Italy\\
\llap{$^p$}Physikalisches Institut, University of Heidelberg, Heidelberg,
           Germany\\
\llap{$^q$}Facolt\`a di Scienze, Universit\`a di Torino, Turin, Italy\\
\llap{$^r$}Bogolyubov Institute for Theoretical Physics, Kiev, Ukraine\\
\llap{$^1$}presently at Physics Department, Yale University, New Haven, USA\\
\llap{$^2$}also Instituto Superior de Tecnolog\`\i as y Ciencias Aplicadas,
           La Habana, Cuba\\
\llap{$^3$}presently at Faculty of Physics and Applied Computer Science, AGH
           University of Science and Technology, Krakow, Poland\\
\llap{$^4$}also Benemerita Universidad Autonoma de Puebla, Puebla, Mexico\\
\llap{$^4$}on leave from Physics Department, Centro de Investigacion y de
           Estudios Avanzados del IPN, Mexico\\
\llap{$^6$}presently at Department of Physics, Creighton University, Omaha,
           NE, USA\\
  E-mail: \email{sitta@to.infn.it}
}

\abstract{The calibration and performance of the Silicon Drift Detector of the
ALICE experiment during the 2008 cosmic ray run will be presented. In
particular the procedures to monitor the running parameters (baselines, noise,
drift speed) are detailed. Other relevant parameters (SOP delay, time-zero,
charge calibration) were also determined.}


\begin{document}

\section{Introduction}

High-energy heavy-ion physics aims to study and attempts to understand how
collective phenomena and macroscopic properties, involving many degrees of
freedom, emerge from the microscopic laws of elemen\-tary-particle physics. The
most interesting case of collective phenomena is the occurrence of phase
transitions in quantum fields at characteristic energy densities; this would
affect the current understanding of both the structure of the Standard Model at
low energy and of the evolution of the early Universe. In ultra-relativistic
heavy-ion collisions, energy densities are expected to reach and exceed the
critical energy density $\epsilon_c \simeq 1$~GeV fm$^{-3}$, predicted by
lattice calculations of Quantum ChromoDynamics (QCD) for a phase transition of
nuclear matter to a deconfined state of quarks and gluons, thus making the QCD
phase transition the only one predicted by the Standard Model that is presently
within reach of laboratory experiments.

ALICE \cite{ppr,techpap} is a general-purpose heavy-ion experiment primarily
designed to study the physics of strongly interacting matter and the
quark--gluon plasma formed in nucleus--nucleus collisions at the LHC
\cite{giubell}. Its detectors measure and identify mid-rapidity ($-0.9 \le \eta
\le 0.9$) hadrons, electrons and photons produced in the collision and
reconstruct particle tracks, including short-lived ones, in an environment with
large multiplicity of charged particles. A forward muon arm detects and
identifies muons covering a large rapidity domain ($-4.0 \le \eta \le -2.4$).
Hadrons, electrons and photons are detected and identified in the central
rapidity region by a complex system of detectors immersed in a moderate (0.5 T)
magnetic field. Tracking relies on a set of high resolution detectors: an Inner
Tracking System (ITS) consisting of six layers of silicon detectors, a
large-volume Time-Projection Chamber (TPC) and a high-granularity
Transition-Radiation Detector (TRD). Particle identification in this central
region is performed by measuring energy loss in the ITS and TPC, transition
radiation in the TRD, Time Of Flight (TOF) with a high-resolution array of
multigap Resistive Plate Chambers, Cherenkov radiation with a High-Momentum
Particle Identification Detector (HMPID), photons with a high granularity
crystal PHOton Spectrometer (PHOS) and a low granularity electromagnetic
calorimeter (EMCAL). Additional detectors located at large rapidities complete
the central detection system to characterize the event and to provide the
interaction triggers.

The Inner Tracking System \cite{techpap,its} is the tracking detector nearest
to the interaction point, and covers the $|\eta| < 0.9$ rapidity range. Its
basic functions are the determination of primary and secondary vertices, the
improvement of the momentum and angular resolution of tracks reconstructed in
the external Time Projection Chamber, as well as the tracking and
identification of particles missed by the TPC.

The ITS consists of six coaxial cylinders: two innermost ones form the Silicon
Pixel Detectors (SPD), two intermediate ones the Silicon Drift Detectors (SDD),
two outermost ones the Silicon Strip Detectors (SSD). The number, position and
segmentation of the layers have been optimized for efficient track finding and
high impact parameter resolution. Due to the enhanced particle density and in
order to achieve the required resolution, different techniques are exploited to
reconstruct in two dimensions the intersection point of a particle with each
layer: in this way the whole ITS allows to precisely reconstruct the trajectory
of particles directly emerging from the interaction point. Moreover the SDD and
SSD also provide $dE/dx$ information needed for ITS particle identification,
while the fast-OR of the SPD read-out chips is used as a Level-0 trigger
signal.

The physics programme of ALICE includes data taking during the pp runs and
dedicated proton-nucleus runs to provide reference data and to address a
specific pp physics programme. Heavy ion running will primarily be done with Pb
ions, but data with lighter ions will also be collected in order to study the
energy-density dependence of the measured phenomena. A prolonged period of
cosmic ray data taking (henceforth called "cosmic runs") is foreseen for the
commissioning phase of various subdetectors; an extended programme of cosmic
runs for physical analyses has also been planned.

In Section 2 the hardware, the front-end electronics and the data acquisition
of the Silicon Drift Detectors are described. In Section 3 the acquisition runs
used to calibrate and monitor the detectors are presented. The analysis of data
taken during the 2008 cosmic run period and their use for detector alignment
and charge calibration are discussed in detail in Section 4. Final conclusions
and future perspectives are addressed in Section 5.

\section{The Silicon Drift Detectors}

The Silicon Drift Detectors (SDD) equip the two intermediate layers (numbered
3 and 4) of the ITS, where the charged particle density is expected to reach 
up to 7~cm$^{-2}$ \cite{techpap}.

One SDD module consists of a drift detector and its front-end electronics. It
was produced from very homogeneous high-resistivity (3~k$\Omega$~cm) 300~$\mu$m
thick Neutron Transmutation Doped (NTD) silicon \cite{sddchar}. It has a
sensitive area of 70.17($r\varphi$) $\times$ 75.26($z$)~mm$^2$ and a total area
of 72.50 $\times$ 87.59~mm$^2$. The sensitive area of a detector is split into
two drift regions, with electrons moving in opposite ($r\varphi$) directions,
by a central cathode kept at a nominal voltage of -1800~V\/. In each drift
region, and on both detector surfaces, 291 p$^+$ cathode strips, with
120~$\mu$m pitch, fully deplete the detector volume and generate a drift field
of $\sim 500$ V/cm parallel to the wafer surface. A second medium voltage (MV)
supply of -40~V keeps the biasing of the collecting region independent of the
drift voltage, and directs the electron charge toward the collecting anodes.

Each drift region is equipped with 256 anodes with 294~$\mu$m pitch (along the
$z$ direction) to collect the charges, and three rows of 33 MOS charge injectors (20 $\times$ 100~$\mu$m$^2$ each) used in monitoring the drift velocity as a
function of the anode number for each SDD module \cite{inject}. This velocity
depends on temperature as $v_{\rm drift} \propto T^{-2.4}$ \cite{vdriftmonit},
giving a 0.8\%/K variation at room temperature. It is about 6.5~$\mu$m/ns at
the bias voltage of $-1.8$~kV and at the normal operating temperature (around
20-25$^\circ$C).

The space precision along the drift direction ($r\varphi$), as obtained during
beam  tests of full-size prototypes, depends on the drift distance and is
better than 38 $\mu$m over the whole detector surface. The precision along the
anode axis ($z$), also dependent on drift distance, is better than 30 $\mu$m
over 94\% of the detector surface and reaches 60~$\mu$m close to the anodes,
where a fraction of clusters affects only one anode. The average values are
35~$\mu$m ($r\varphi$) and 25~$\mu$m ($z$) respectively \cite{nouais}.

\begin{figure}[ht]
\begin{center}
\includegraphics[scale=0.172]{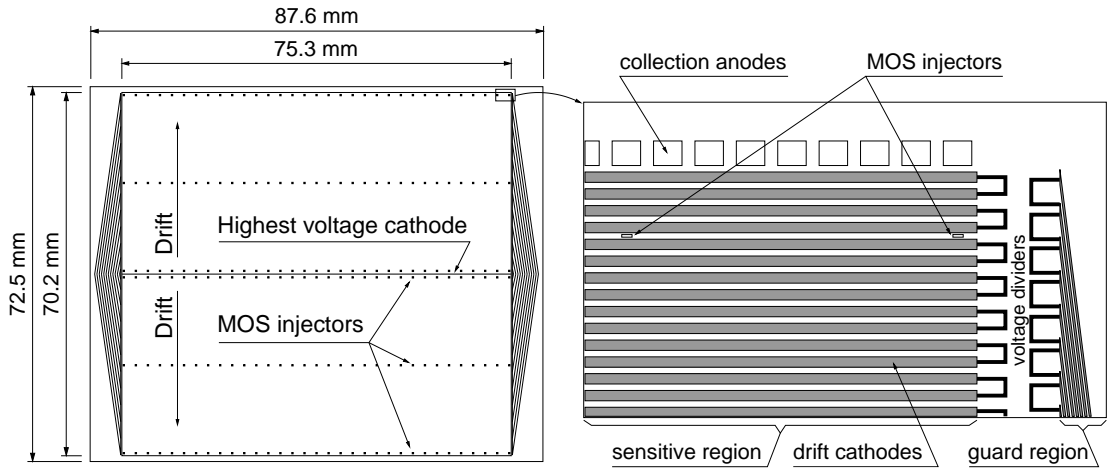}
\end{center}
\caption{Layout of the ALICE SDD modules. The left panel shows a whole module,
the right panel is an enlargement of the collecting anode region.
}
\label{fig:sddlayout}
\end{figure}

The main characteristics of each detector are summarized in Table
\ref{tab:sdd}; a sketch of one sensor is depicted in Figure \ref{fig:sddlayout}

\begin{table}[t]
\caption{SDD main characteristics.\label{tab:sdd}}
\vspace{-0.3cm}
\begin{center}
\begin{tabular*}{0.485\textwidth}{@{\extracolsep{\fill}} lc|lc}
\multicolumn{3}{c}{Sensitive area}{$70.17\times75.26$~mm$^2$}\\
HV (nominal) & -1800 V & Bias (MV) & -40 V\\
Anode pitch & 294 $\mu$m & Drift velocity & $\sim$6.5 $\mu$m/ns\\
Av. $z$ resolution & 25 $\mu$m & Av. $r\varphi$ resolution & 35 $\mu$m\\
\end{tabular*}
\end{center}
\end{table}

The SDD modules are mounted on linear support structures called {\it
ladders}\/: on layer 3 (at $r = 14.9$ cm) there are 14 ladders with 6 modules
each, while on layer 4 (at $r = 23.8$ cm) there are 22 ladders with 8 modules
each, for a total of 260 modules. To ensure full angular coverage ladders and
modules are assembled with an overlap of the sensitive areas larger than
580~$\mu$m in both $r\varphi$ and $z$ directions (corresponding to about 2\% of
the module area). During the 2008 cosmic run $\sim$~98\% of the SDD modules
were included in the acquisition.

During the module assembling on ladders, the position of each module was
precisely measured with respect to a fixed point (a ruby sphere fixed to the
ladder structure). The resulting residual between the real and nominal
position is centered on 0 and has an RMS of about 20 $\mu$m for both $z$ and
$r\varphi$ directions. 

All 260 modules were characterized before being assembled in ladders using an
infrared laser. Charges were injected in $>100,000$ positions per module, and
residuals between the indirectly reconstructed coordinates (from measured ADC
counts) and the known laser ones were computed. In this way a map of systematic
deviations of drift coordinate (mainly caused by non-constant drift field
influenced by voltage supplied by non-linear voltage dividers and in two cases
by parasitic electric fields due to significant dopant inhomogeneities) was
constructed for each detector, to be used to correct the time coordinate during
the reconstruction stage \cite{dopant}.

\subsection{Front-end electronics and readout}
\label{subsecFERO}

\begin{figure}[t!]
\begin{center}
\begin{tabular}{c c}
\mbox{\includegraphics[width=0.44\linewidth]{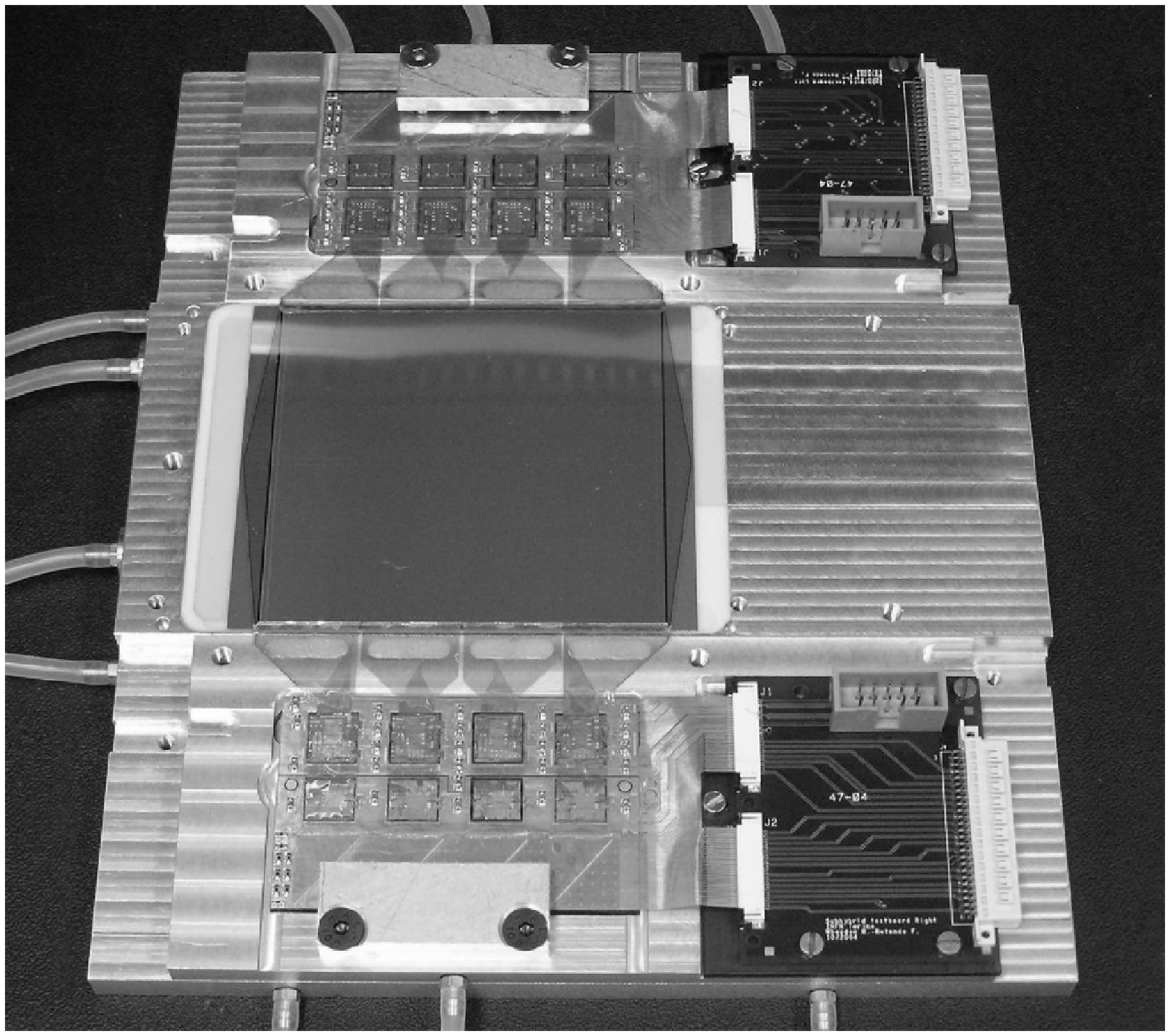}}
&
\mbox{\includegraphics[width=0.50\linewidth]{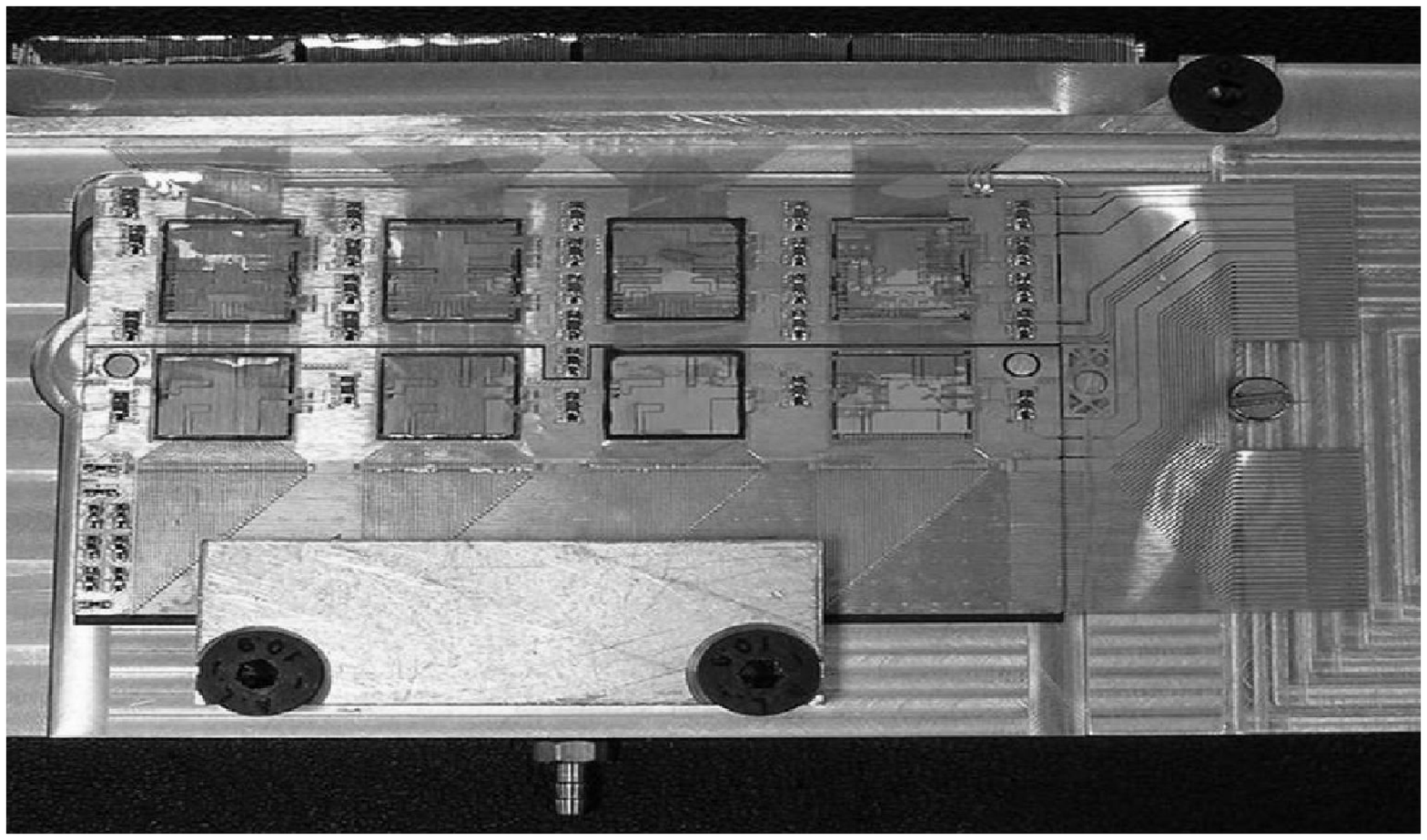}}
\end{tabular}
\end{center}
\caption{Left: A Silicon Drift Detector on the assembly jig with the two
front-end hybrids connected, via their test extensions, to test interface
circuits. Right: A single hybrid on its assembly jig: the four chips on 
the top row are the PASCALs, the ones on the bottom row are the AMBRAs.}
\label{modass}
\end{figure}

The SDD front-end electronics is based on three types of ASICs, two of them,
PASCAL and AMBRA, assembled on an hybrid circuit (shown in Figure \ref{modass}
on its assembly jig) which is directly bonded to the sensor, and one, CARLOS,
located at each end of a ladder.

Each PASCAL chip, which reads signals from 64 anodes, contains three functional
blocks: preamplifier, analogue storage and  Analogue-to-Digital Converter
(ADC). Each AMBRA chip receives the input from a PASCAL, and is a digital
four-event buffer which performs data derandomisation, baseline equalization
on an anode-by-anode basis and 10-bit to 8-bit non-linear data compression. The
output of an AMBRA is fed to a CARLOS chip, which performs data compression and
bidimensional two-threshold zero suppression. The nominal power supplies are
2.5~V for each PASCAL, AMBRA and CARLOS respectively. The average power
dissipation of each PASCAL--AMBRA front-end channel is about 6~mW. A cooling
system is used to take away the power dissipated by the front-end electronics
and to maintain a temperature stability of the order of 0.1~K \cite{cool}. The
system uses demineralized water as cooling agent and is divided into 52
independent circuits. In order to provide leak-tightness inside the detector
the water pressure is kept below the atmospheric one. This condition is
constantly monitored by a dedicated interlock system. Moreover the water flow
is measured in every circuit in order to assure the adequate power absorption.

The signal generated by an SDD anode feeds the PASCAL amplifier-shaper which
has a  peaking time of 40 ns and a dynamic range of 32 fC (corresponding to the
charge released by an 8-MIP particle hitting near the anode). The amplifier
output is sampled at 40 MHz by a ring analogue memory with 256 cells per anode.
On a run-by-run basis, PASCAL chips can be programmed to use half of the 40 MHz
nominal frequency for the sampling, thus reducing the amount of data and,
therefore, the sub-system readout dead-time. When a Level0 (L0) trigger is
received, the SDD BUSY is immediately set and after a programmable delay which
accounts for the L0 latency (1.2~$\mu$s) and the maximum detector drift time
($\sim$5~$\mu$s) the analogue memories are frozen. The BUSY being still set,
their contents are then digitised by a set of 10-bit linear
successive-approximation ADCs which write the data into one of the free AMBRA
buffers. The digitisation can be aborted by the absence of the Level1 trigger
or by the arrival of a Level2-reject signal: in both cases, the front-end
electronics resets the SDD BUSY and returns to the idle state. On the
successful completion of the analogue-to-digital conversion the SDD BUSY is
reset if at least one buffer is still available in the AMBRAs.

As soon as the conversion is completed, all the AMBRAs transmit the data in
sequence to the CARLOS chips. The zero suppression is performed by a
2-dimensional 2-thresholds algorithm: to be accepted as a valid signal, the
ADC counts in a memory cell must exceed the higher threshold and have at least
one neighbour above the lower one or viceversa. This allows to suppress noise
spikes and to preserve as much as possible the samples in the tail of the
signals \cite{twothres}. With no additional dead time, using this algorithm
the CARLOS chips reduce the SDD event size by more than one order of magnitude.
They also format the data and feed the Gigabit Optical Link (GOL) ASICs which
in turn send the data to the DAQ electronics via an optical fiber (about 100~m
long).

The three ASICs embed a JTAG standard interface, which is used to download
control information into the chips before the data taking. By means of the same
interface it was possible to test each chip after the various assembly stages.
A programmable test pulse generator on PASCAL allows a fast, yet detailed, test
of the whole chain for calibration purposes.

\subsection{Data acquisition}

In the counting room, 24 VME boards, CARLOSrx \cite{carlosrx}, concentrate the
data coming from the 260 SDDs into 24 Detector Data Link (DDL) channels and
embody the trigger information in the data flow. Each CARLOSrx board controls
up to 12 CARLOS chips. CARLOSrx also uploads in parallel the configuration
parameters received over the DDL to the ladder front-end electronics it
controls and monitors the error-flag words embedded in the data flow coming
from the CARLOS chips in order to signal potential Single-Event Upsets (SEU) on
the ladder electronics.

The CARLOSrx boards are arranged in three VME crates, one devoted to the
acquisition of all modules on layer 3 and the other two for the modules of
layer 4. Each VME crate is equipped with a VME CPU card used to load the
board's firmware; this operation can be performed remotely by the operator from
the experiment's control room.

Each CARLOSrx board is connected via an optical link to one input of a D-RORC
(DAQ ReadOut Receiver Card) \cite{drorc} PCI card (the D-RORC has two outputs
and sends an exact copy of the data both to the DAQ and to the High-Level
Trigger system). Since a DAQ PC can host a maximum of six D-RORC cards, during
the 2008 cosmic data taking period presented here four Local Data Concentrator
(LDC) computers were used to manage the SDD acquisition. Each LDC interfaces
with the main run control programs, performs data collection and sub-event
building, stores data locally and transfers them to the Global Data Collector
(GDC) nodes if requested. During 2008 one GDC was dedicated to SDD standalone
data taking. The GDC collects sub-events from the LDCs, performs the event
building and stores data either locally or on a mass-storage system. Data
recording can be disabled by software both on the LDC and GDC level. LDC's
allow also to test the uploading of JTAG commands on the SDD front-end chips
and to dump the data flow to a terminal for test and monitoring purposes.

When running in standalone mode the whole data taking process can be controlled
by the SDD operator. In standalone mode the detector can acquire data using a
pulsed trigger or a random trigger (whose rate can be set from 1 Hz up to 40
MHz). This allows to perform the calibration runs described below, to check the
acquisition operation and to verify the detector performance (maximum event
rate, detector occupancy due to noise, etc.). On the other hand when being part
of a global acquisition run, the control is locked by the main Experiment
Control System (ECS) and the SDD operator can only monitor the data flow and
quality.

The SDD online monitoring is based on the Quality Assurance (QA) tools that are
part of the ALICE offline framework, so as to provide online and offline
functionality without code duplication. It is interfaced to the ALICE DAQ Data
Quality Monitoring (DQM) framework AMORE (Automatic MOnitoRing Environment)
\cite{barth}, and is based on a publisher-subscriber approach. The SDD QA code
is driven by an AMORE Agent that publishes in the AMORE Database the histograms
generated by the SDD QA code itself. These histograms include the pattern of
fired ladders and modules for each layer, and the charge map distributions of
each half module (for a sampled event and integrated on all the analysed
events). All the distributions can be retrieved from the database and
visualized with the help of a custom Graphical User Interface (SDD GUI) that
works as subscribing client. Distributions are plotted in groups selected by
specific tabs. The currently available tabs provide the summary plots of the
module pattern of each ladder for each layer, the DDL connections, the
projection of the 2-D charge maps and the module maps.

\section{Detector Calibration}

Three types of calibration runs have been implemented to monitor the
performance of the detectors and to extract the relevant parameters for event
reconstruction. These runs are initiated by the SDD operator. Events are
triggered by the Local Trigger Unit (LTU) whose trigger rate is manually set to
a very low frequency (not more than 4 Hz). After a predefined number of events
is collected, the run is automatically stopped. The collected data are
registered on both the LDCs (for immediate analysis) and on the mass storage
system (for possible offline re-analysis). A dedicated Detector Algorithm (DA)
is executed on the LDCs immediately after the end of a calibration run to
analyze these data and calculate the calibration parameters that are needed in
the reconstruction phase. These calibration quantities provided by the DAs are
stored in the Offline Condition DataBase (OCDB) for the SDD by means of the
SHUTTLE framework \cite{shuttle}. The reconstruction code can then retrieve
this information during the offline reconstruction by automatically selecting
the most recent set of calibration data according to the run being analyzed.

During 2008 these runs were carried out once a day, when the detector was
temporarily not part of a global run. For the future it is planned to run these
calibrations during the LHC fill periods (every $\approx$ 24 hours).

\subsection{Pedestal runs}

Pedestal runs are special standalone runs performed every $\approx$ 24 hours
with both zero suppression and baseline equalization disabled. Their analysis
provides a measurement of the anode baseline (defined as the average signal
value in ADC counts for each anode) and of the noise (defined as the standard
deviation of the above determination of the average for each anode) both raw
and corrected for common mode. Moreover they can tag the noisy anodes which can
be masked out during physics runs. The values of baselines and noise and the
fraction of bad channels (around 0.5\%) have all proven to be very stable
during this entire data taking period.

When starting a pedestal run proper JTAG commands are sent to all modules to
inhibit baseline equalization and set to zero the thresholds for zero
suppression. The results of the pedestal DAs are the computation of the anode
baselines and noise.

\begin{figure}[t!]
\begin{center}
\begin{tabular}{c c}
\mbox{\includegraphics[width=0.44\linewidth]{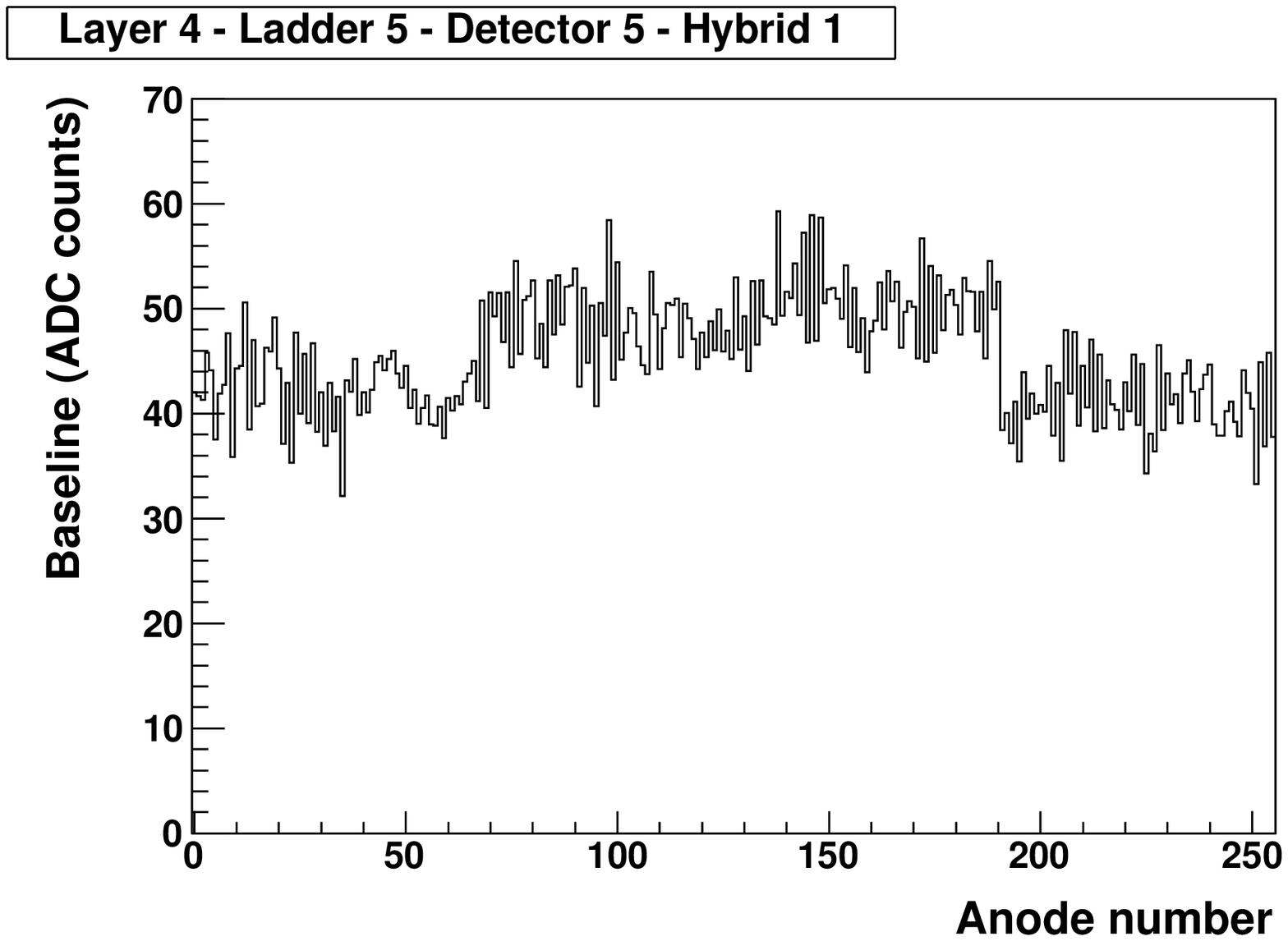}}
&
\mbox{\includegraphics[width=0.44\linewidth]{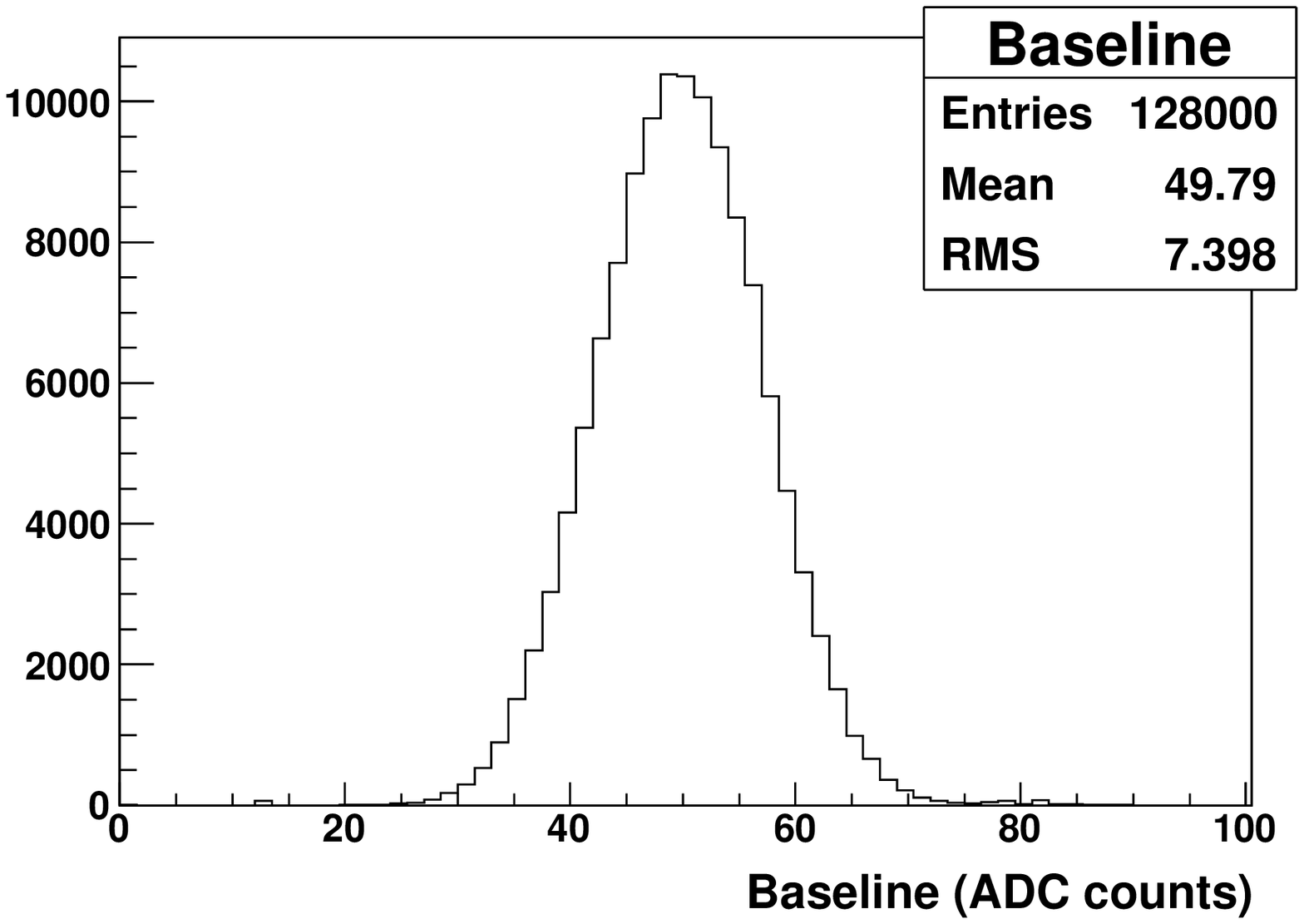}}
\end{tabular}
\end{center}
\caption{Left: Baseline distribution as a function of the anode number of an
half module. Right: Distributions of all baseline values before equalization in
a pedestal run.}
\label{baseline}
\end{figure}

An example of baseline distribution as a function of the anode number for one
half module is shown in Figure \ref{baseline} left; on the right panel of the
same Figure there is the distribution of baselines before equalization for all
modules. Baselines are then equalized to
bring them all to the same level of 20; the equalization value for each anode
is inserted in the JTAG command sequence loaded on the AMBRA chips. The value
of 20 was chosen as a compromise (to allow the 6-bit correction of high
baseline anodes to be sufficient and to have a common baseline for all modules)
between different requests: the maximum equalization correction is 6 bits (63
ADC units); the baseline should be as low as possible to take the best
advantage of the 0-127 count range in which the ADC is not affected by the
non-linear compression from 10 to 8 bits; on the other hand the baseline should
not be so low that the noise can bring the ADC counts below zero.

The same pedestal run is used to compute the noise, both raw and corrected for
common mode, and to tag the noisy anodes. Anodes are marked as bad when the raw
noise is $< 0.5$ or $> 9$ ADC counts; moreover anodes whose noise is greater
than 4 times the mean noise of the entire module are also tagged as noisy.
These bad anodes are masked out by inserting a proper bit mask in the JTAG
commands sent to the AMBRA chips.

\begin{figure}[ht]
\begin{center}
\begin{tabular}{c c}
\mbox{\includegraphics[width=0.44\linewidth]{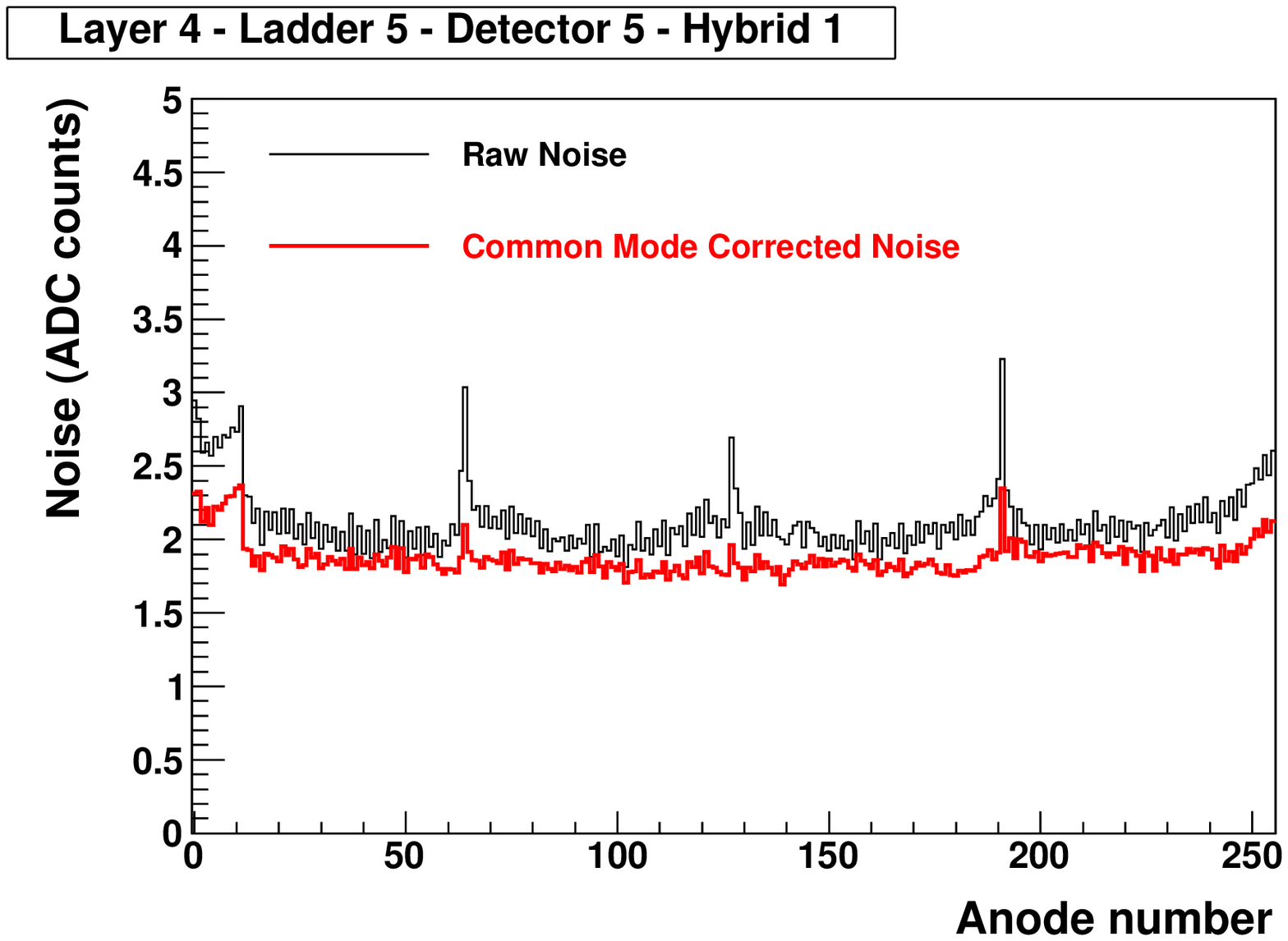}}
&
\mbox{\includegraphics[width=0.44\linewidth]{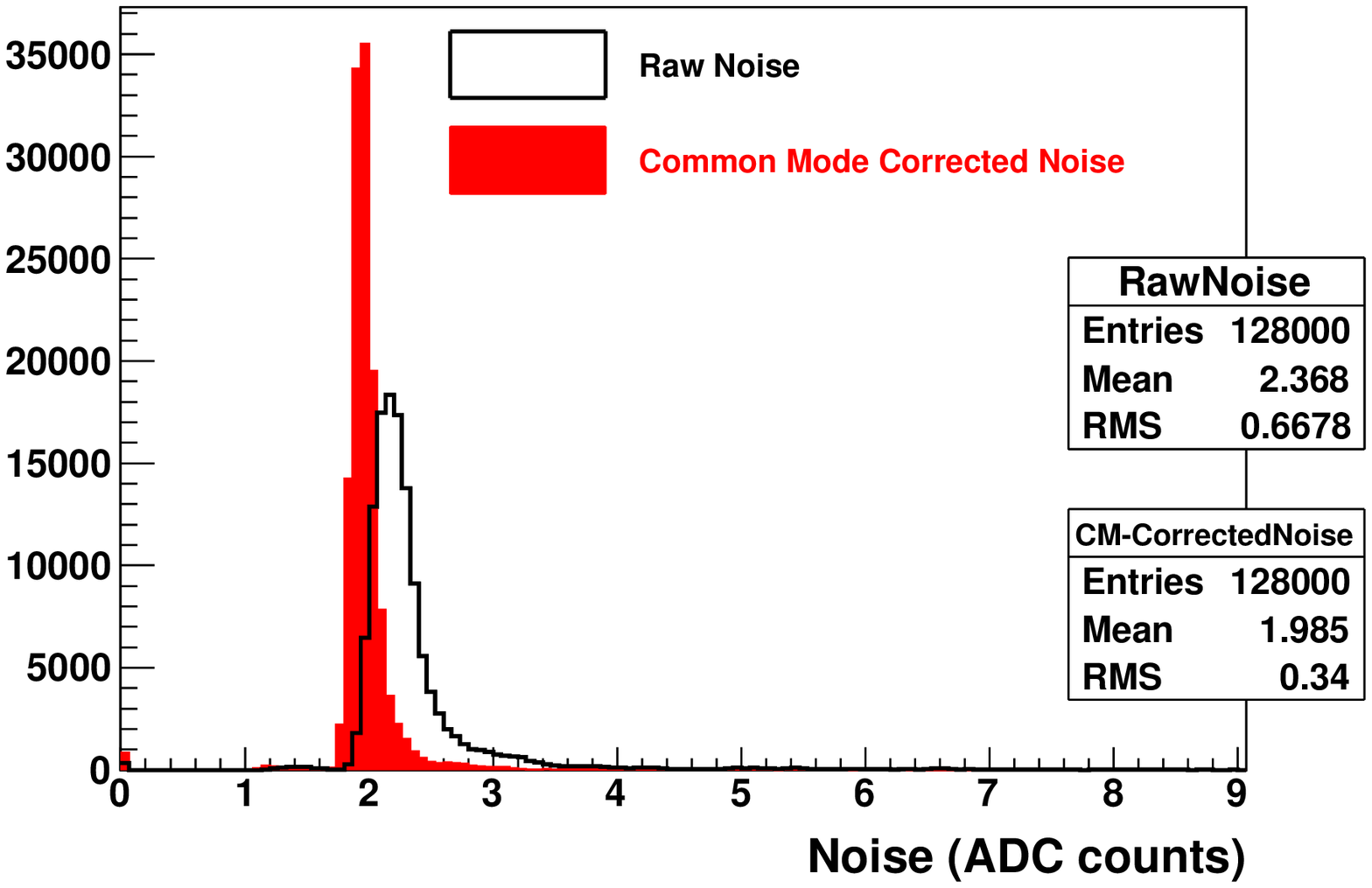}}
\end{tabular}
\end{center}
\caption{Left: Raw (black) and common mode corrected (red) noise distributions
as a function of the anode number of an half module. Right: Distributions of
raw (black) and common mode corrected (red) noise values for all anodes.}
\label{noise}
\end{figure}

In Figure \ref{noise} (left panel) the noise distribution as a function of the
anode number of one half module is shown, while on the right panel of the same
Figure are the distributions of the raw and common mode corrected noise for all
anodes. The peaks in the raw noise are located around anodes 0, 64, 128, 192
and 255, that is on the chip borders, since these anodes are more sensitive to
the common mode noise.

The measured noise is in agreement with the design value of 2 ADC counts,
corresponding to an Equivalent Noise Charge (ENC) of 350 electrons. Due to the
diffusion effect, the gaussian width of the electron cloud grows with the
square root of the drift time, thus giving raise to a drift time dependent
cluster size. For a particle hitting close to the anodes a typical cluster size
is 1 anode times 3 time bins (at 40 MHz sampling rate), which increases to 3
anodes times 7 time bins for particles with maximum drift distance. The signal
peak (i.e.\ the cell with the maximum ADC counts in the cluster) is typically
of 100 ADC counts over the baseline near the anodes (corresponding to a S/N
ratio of about 50) and decreases with increasing drift distance. The low and
high thresholds for zero suppression are set for each hybrid to 2.2 and 4 times
the average noise respectively.

\begin{figure}[ht]
\begin{center}
\begin{tabular}{c c}
\mbox{\includegraphics[width=0.47\linewidth]{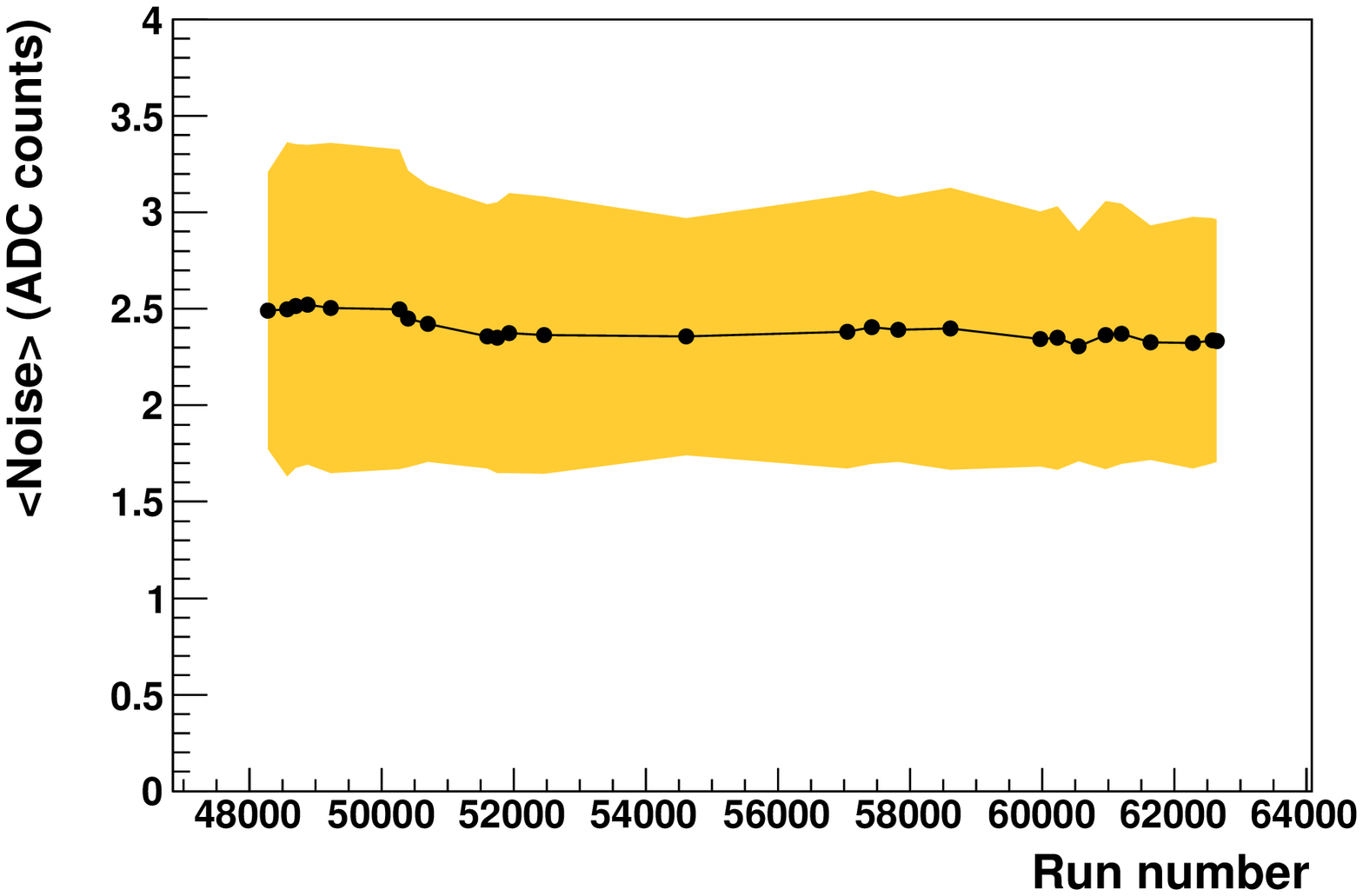}}
&
\mbox{\includegraphics[width=0.47\linewidth]{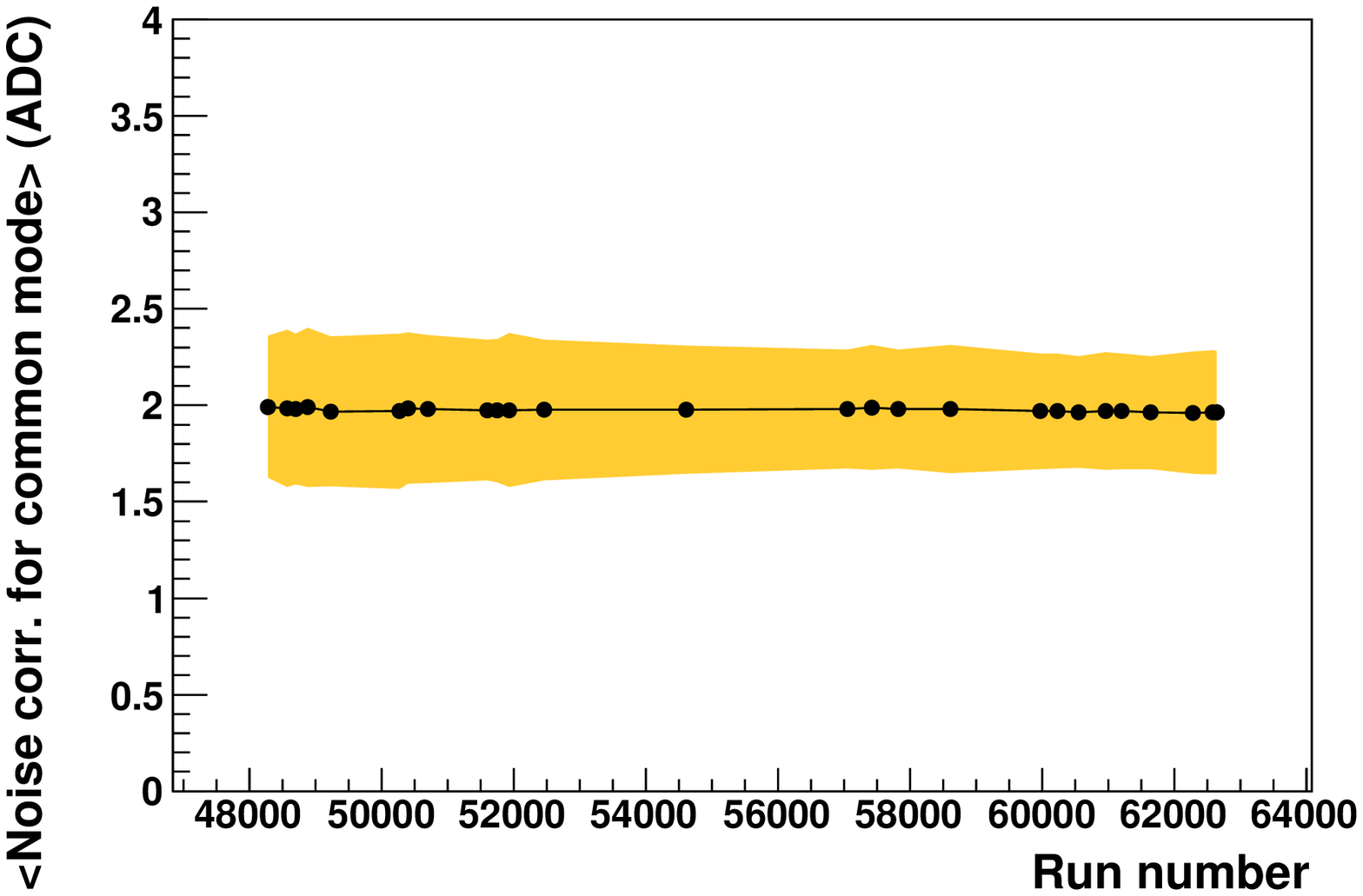}}
\end{tabular}
\end{center}
\caption{Raw (left) and common mode corrected (right) noise trend as a function
of the run number during the 2008 cosmic run (from mid July to mid October);
the yellow band is the RMS of the noise distributions.}
\label{noistime}
\end{figure}

\begin{figure}
\begin{center}
\begin{tabular}{c c}
\mbox{\includegraphics[width=0.47\linewidth]{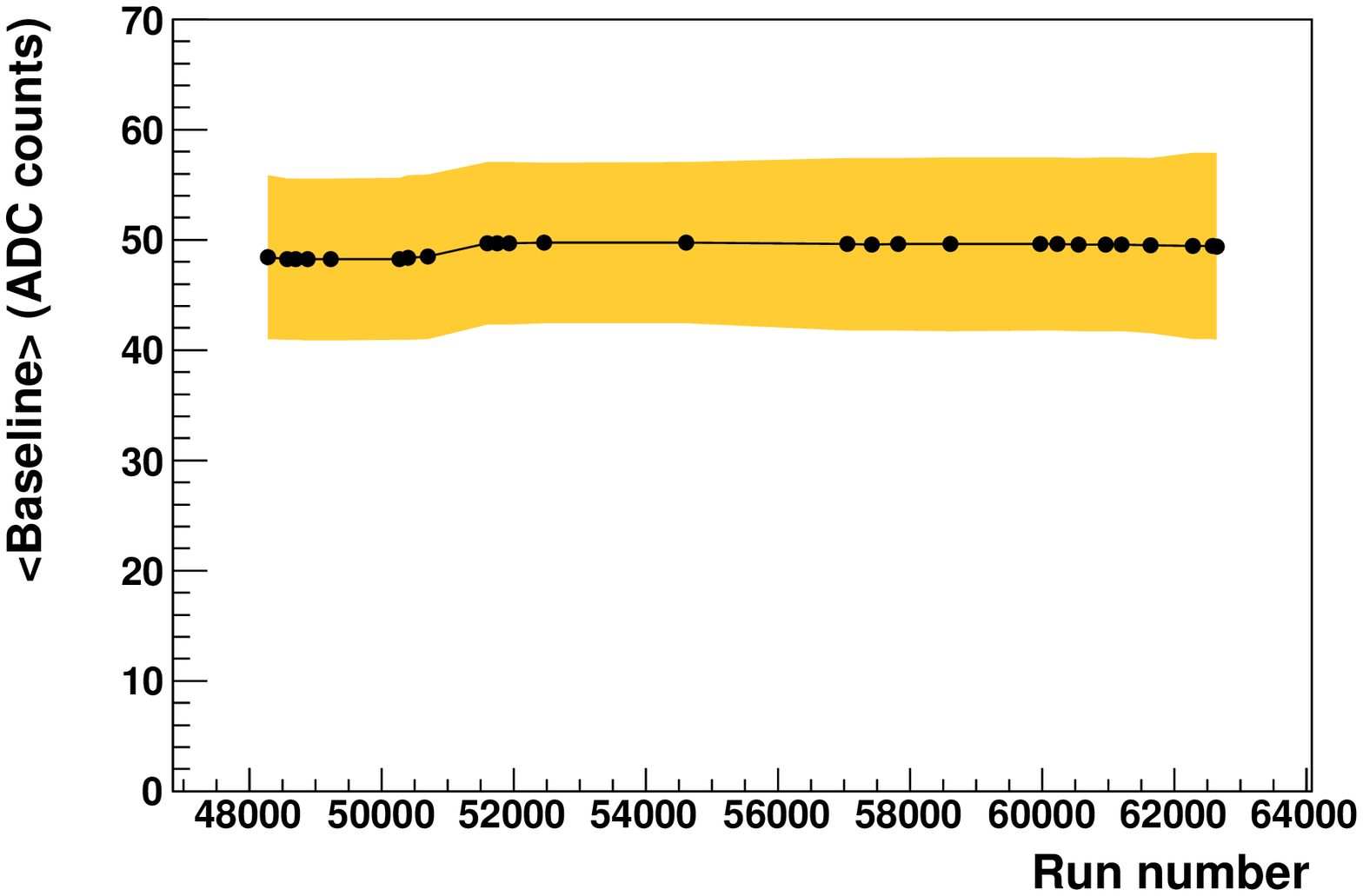}}
&
\mbox{\includegraphics[width=0.47\linewidth]{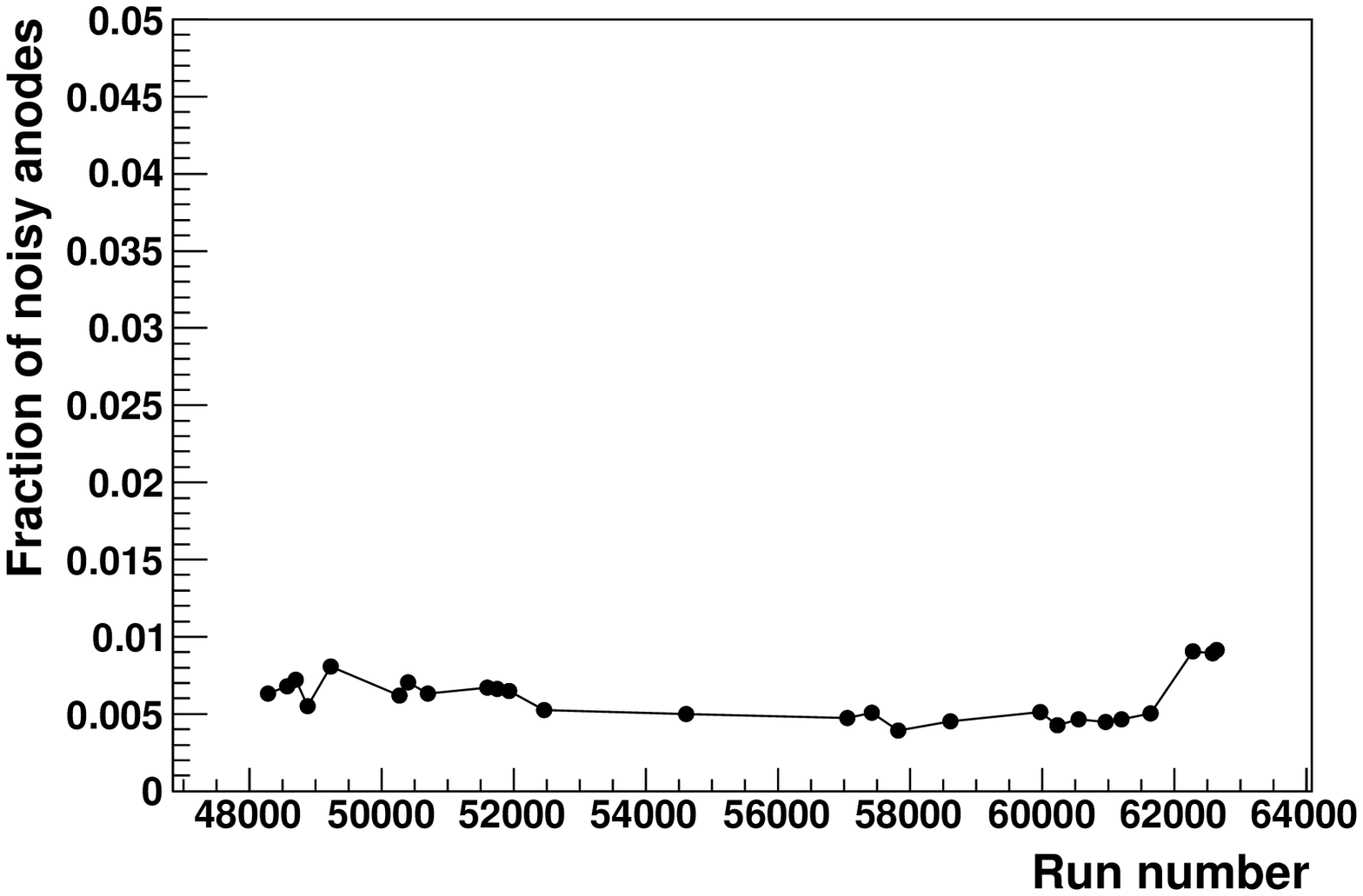}}
\end{tabular}
\end{center}
\caption{Trend of the baseline values (left) and of the fraction of bad anodes
(right) as a function of the run number during the 2008 cosmic run (from mid
July to mid October); the yellow band is the RMS of the baseline distribution.}
\label{basbad}
\end{figure}

The monitored parameters were very stable over a long period of time. In Figure
\ref{noistime} the peak value of the noise distributions, both raw (left) and
corrected for the common mode (right), is reported as a function of the run
number from July 23rd to October 16th (the band is the RMS of each
distributions). Similarly the mean value of the baseline distribution (Figure
\ref{basbad} left) is also stable and the fraction of bad anodes (right) stays
below 1\% over the same period of time.

\subsection{Pulser runs}

Pulser runs are special standalone runs performed every $\approx$ 24 hours and
collected without zero suppression, without baseline equalization and by
sending the test pulse signal to the input of the PASCAL pre-amplifiers. One
test pulse per event is used with amplitude of 100 DAC units, which is slightly
lower than that of a MIP particle (which corresponds to 111 units). The
analysis of a pulser run provides a measurement of the preamplifier gain, and
can tag the dead channels.

A pulser run is normally executed right after a pedestal run. Proper JTAG
commands are sent to all modules to inhibit the zero suppression, set to zero
the thresholds and the baseline equalization, and activate the PASCAL test
pulse generator, so that for each anode the test pulse is sent to the input of
the front-end electronics. The result of pulser DAs is the determination of the
gain for each channel. The information coming from the last pedestal run and
the current pulser run are combined together and saved in the OCDB for the SDD.

\begin{figure}[t]
\begin{center}
\begin{tabular}{c c}
\mbox{\includegraphics[width=0.44\linewidth]{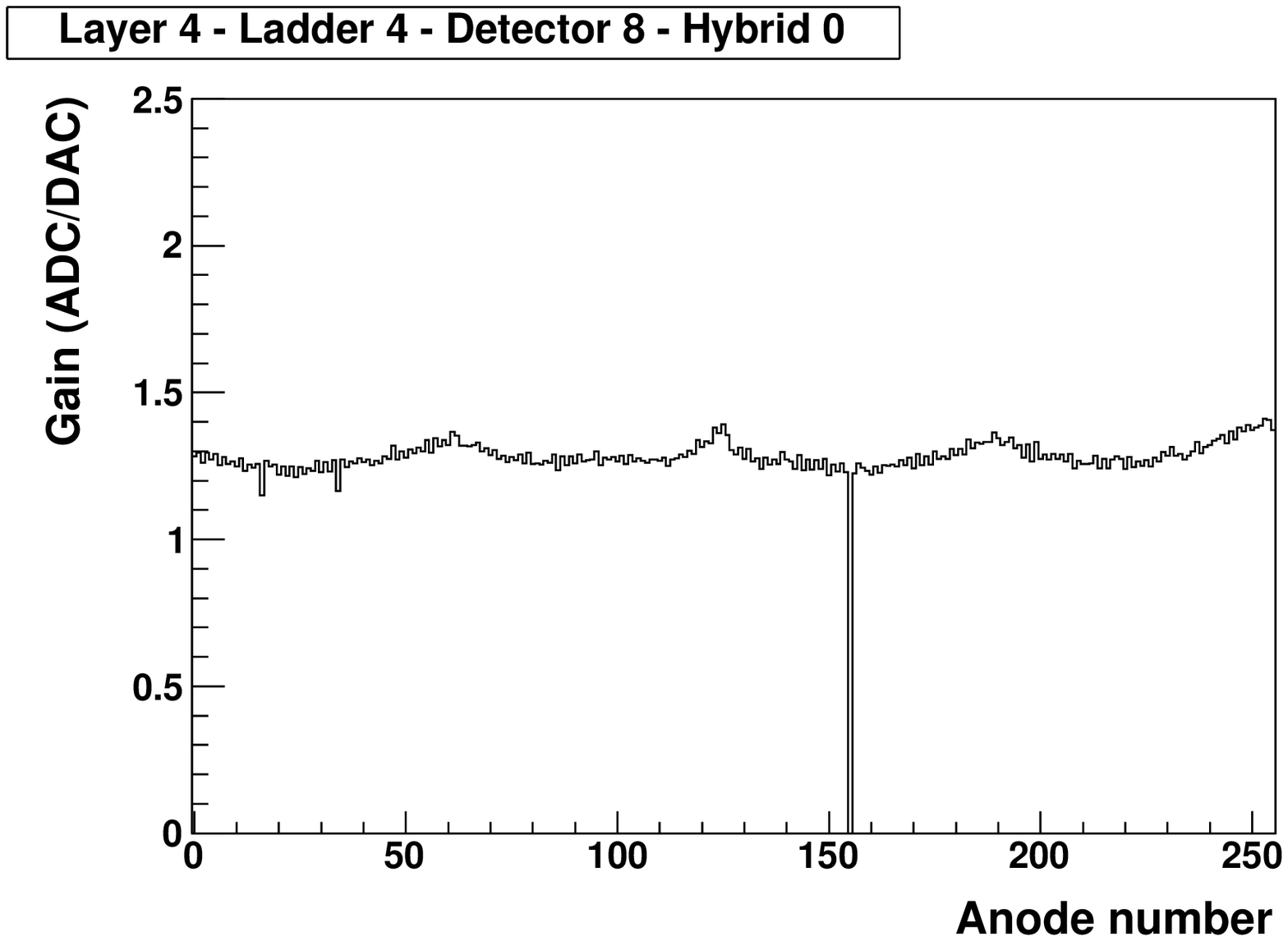}}
&
\mbox{\includegraphics[width=0.44\linewidth]{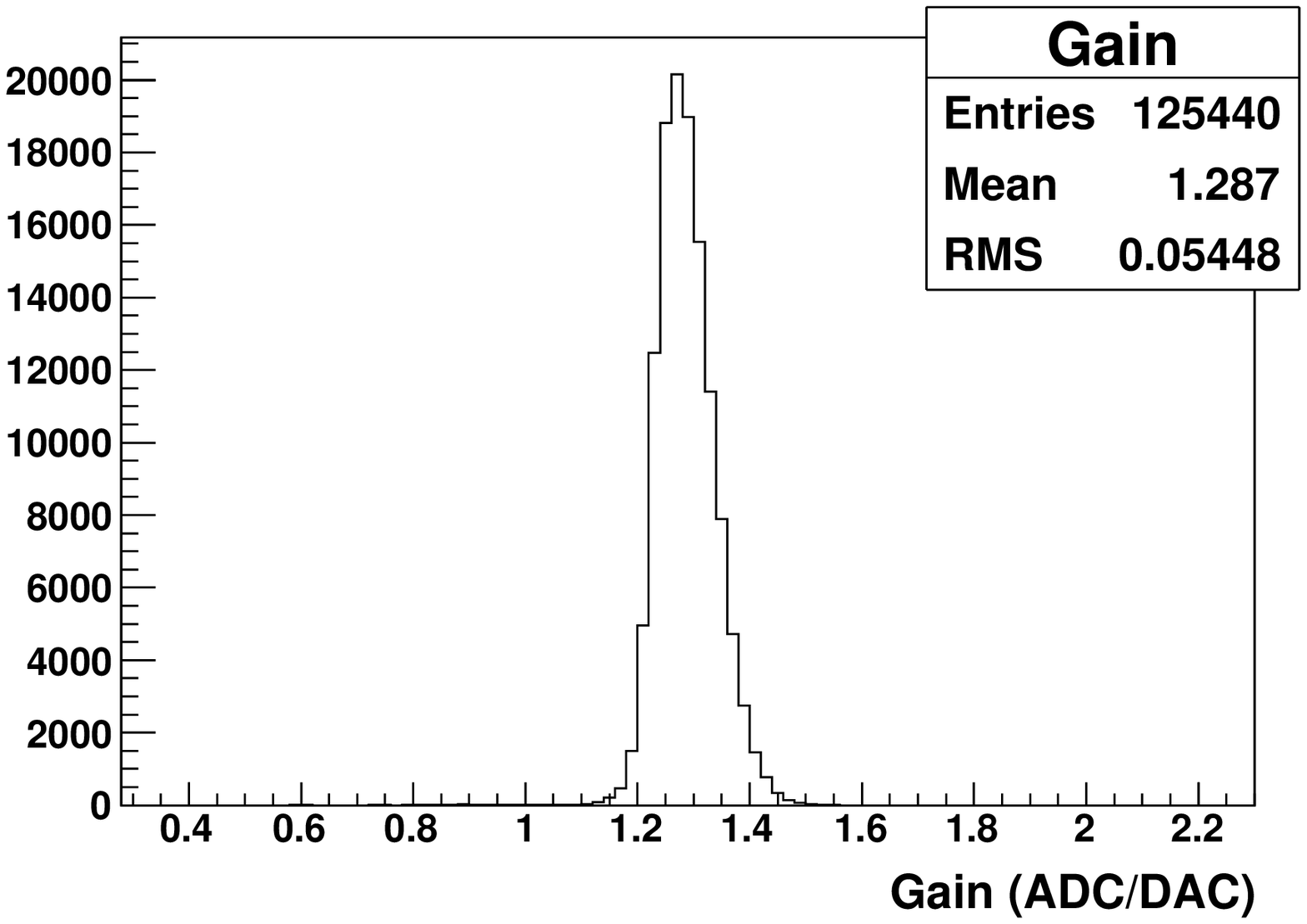}}
\end{tabular}
\end{center}
\caption{Left: Gain distribution as a function of the anode number of an
half module; the bin with zero content corresponds to a dead channel. Right:
Distributions of all gain values in a pulser run.}
\label{gain}
\end{figure}

An example of gain distribution as a function of the anode number of one half
module is shown in Figure \ref{gain} left; the bin with zero content
corresponds to a dead channel. On the right panel of the same Figure \ref{gain}
there is the distribution of gains for all modules. This distribution is used
to equalize the anode gain in the cluster finder algorithm; this correction is
anyway very small since the gain distribution is very narrow.

\begin{figure}[t]
\begin{center}
\begin{tabular}{c c}
\mbox{\includegraphics[width=0.44\linewidth]{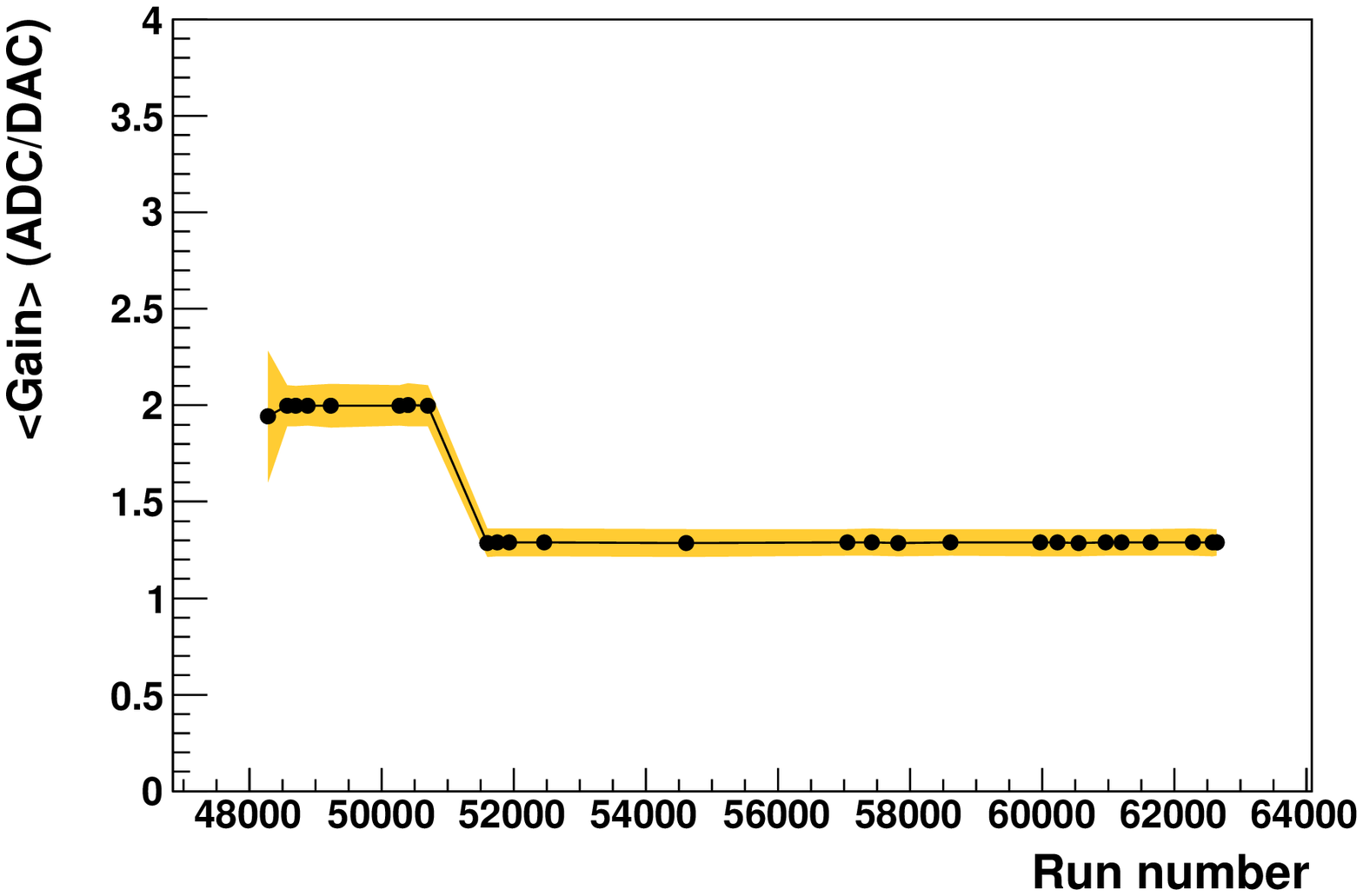}}
&
\mbox{\includegraphics[width=0.44\linewidth]{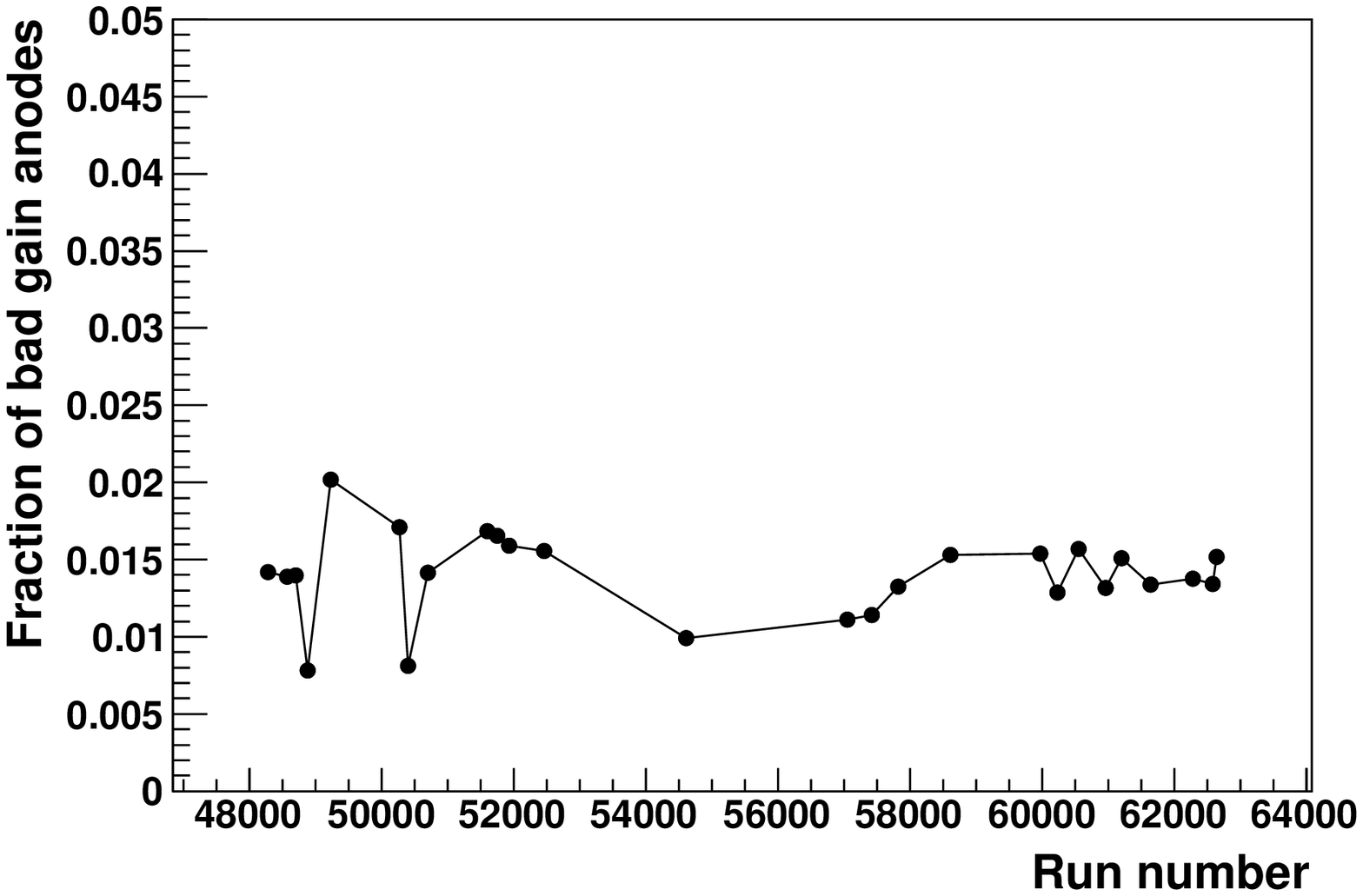}}
\end{tabular}
\end{center}
\caption{Trend of the gain values (left) and of the fraction of dead channels
(right) as a function of the run number during the 2008 cosmic run (from mid
July to mid October); the grey band is the RMS of the gain distribution. The
sudden change of gain value corresponds to the change from 40 MHz to 20 MHz
sampling rate.}
\label{gaintime}
\end{figure}

The monitored parameters were very stable over a long period of time. In the
left panel of Figure \ref{gaintime} the mean value of the gain distribution is
shown as a function of the run number from run July 23rd to October 16th (the
band is the RMS of the distribution). The sudden reduction of gain value
corresponds to the change from 40 MHz to 20 MHz sampling rate of the PASCAL
amplifier; this reduction is connected to the different number of samplings
taken on the impulse response of the front-end pre-amplifier, whose duration is
about 100 ns. In the right panel of the same Figure the fraction of dead
channels stays below 2\% during the same period of time. The fluctuations are
also due to the fact that the sample of modules in acquisition was not constant
during the data taking period.

\subsection{Injector runs}

A precise knowledge of the drift speed is a crucial element for the correct
operation of any drift detector. Given its strict dependence on the detector
temperature, the drift speed must be very frequently measured and monitored. To
reach the design precision of 35 $\mu$m on a drift distance of as much as 35 mm
(from the point farthest from the anodes) the drift speed must be known with an
accuracy of better than 0.1\%\/.

Injector runs are special standalone runs with zero suppression and with
baseline equalization used to measure the drift speed. The MOS injectors are
activated in order to inject charges in known positions (see Fig.\
\ref{fig:sddlayout}).

In 2008 injector runs were performed every $\approx$ 6 hours. Proper JTAG
commands are sent to all modules to enable the zero suppression, equalize the
baselines and activate the MOS charge injectors.  The result of the injectors
DAs is the computation of the drift speed as a function of anode number for
each SDD half-module. This information also is saved in the OCDB for the SDD
subdetector to be used during reconstruction.

\begin{figure}[t]
\begin{center}
\begin{tabular}{c c}
\mbox{\includegraphics[width=0.46\linewidth]{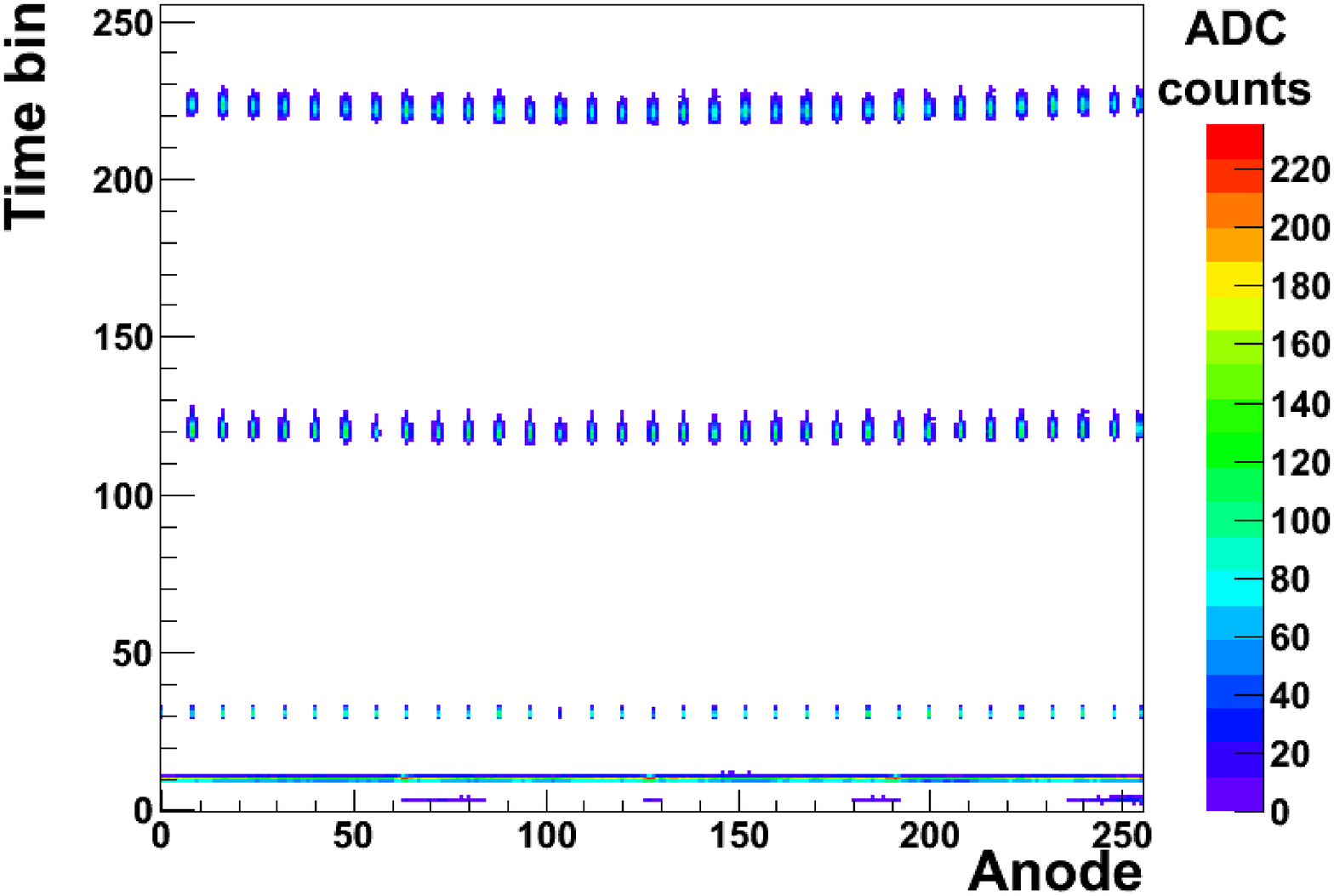}}
&
\mbox{\includegraphics[width=0.46\linewidth]{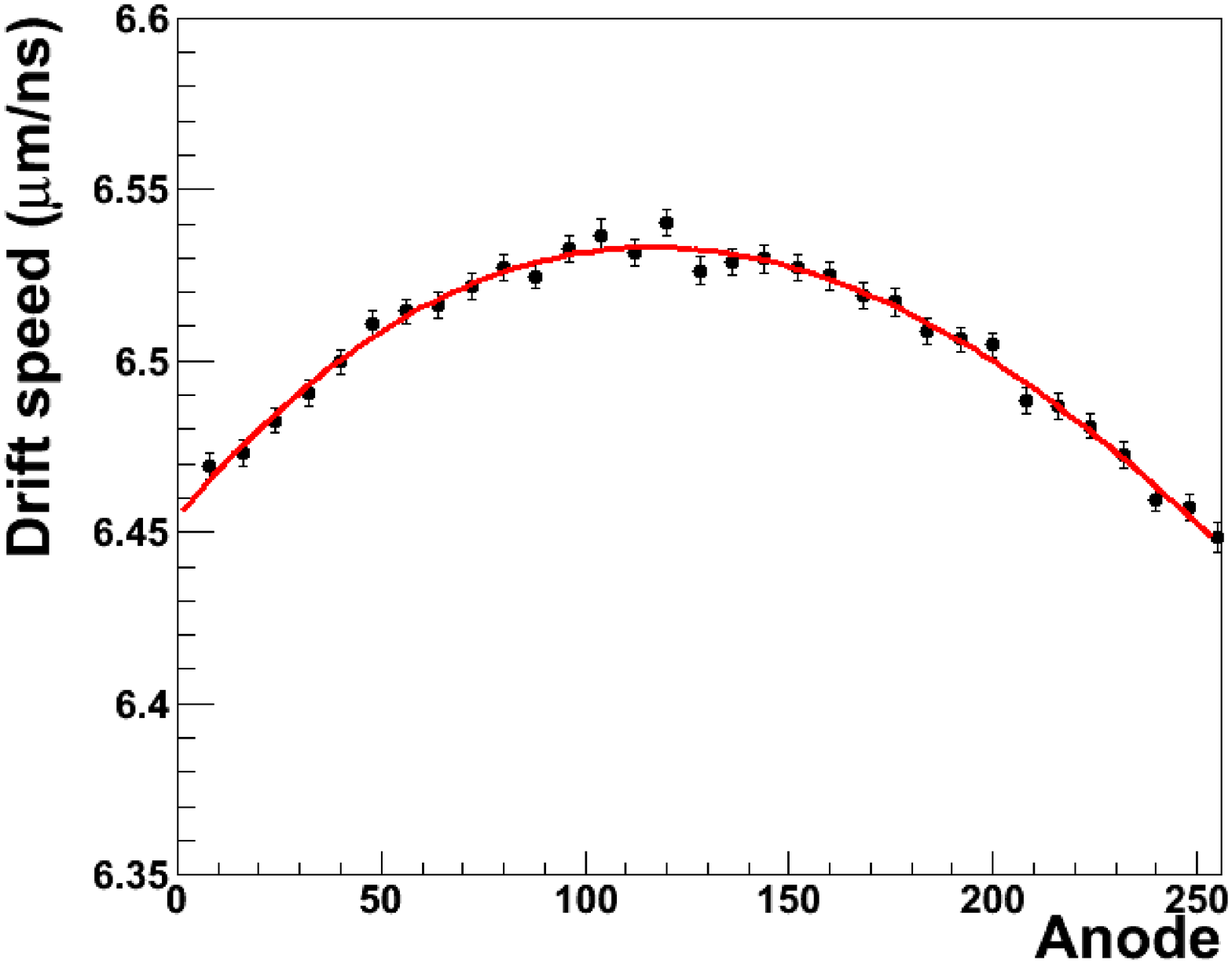}}
\end{tabular}
\end{center}
\caption{Left: Display of an injector event for a half module: the three MOS
injector lines are clearly visible (the horizontal axis reports the anode
number, the vertical axis the time bin; the color chart represents the quantity
of charge collected by a single anode). Right: Drift values as a function of
the anode number of one half module; each point represents the result of the
fit on the three corresponding injectors.}
\label{vdrift}
\end{figure}

In the left panel of Figure \ref{vdrift} the display of one injector event for
one half module is shown: the image of the
three injector lines are clearly visible, together with the pulse indicating
the trigger time. For each anode in front of which there are the MOS injectors
(one anode out of eight) a linear fit of the three measured drift times is
performed as a function of the known drift distance; the drift speed is then
extracted from the slope of the fit. The operation is repeated on all the
injector triplets of each module. An example of drift values as a function of
the anode number of one half module is reported in the right panel of Figure
\ref{vdrift}: the plot shows 33 drift values, one per MOS injector triplet,
with a polynomial fit superimposed.

\begin{figure}[t]
\begin{center}
\begin{tabular}{c c}
\mbox{\includegraphics[width=0.48\linewidth]{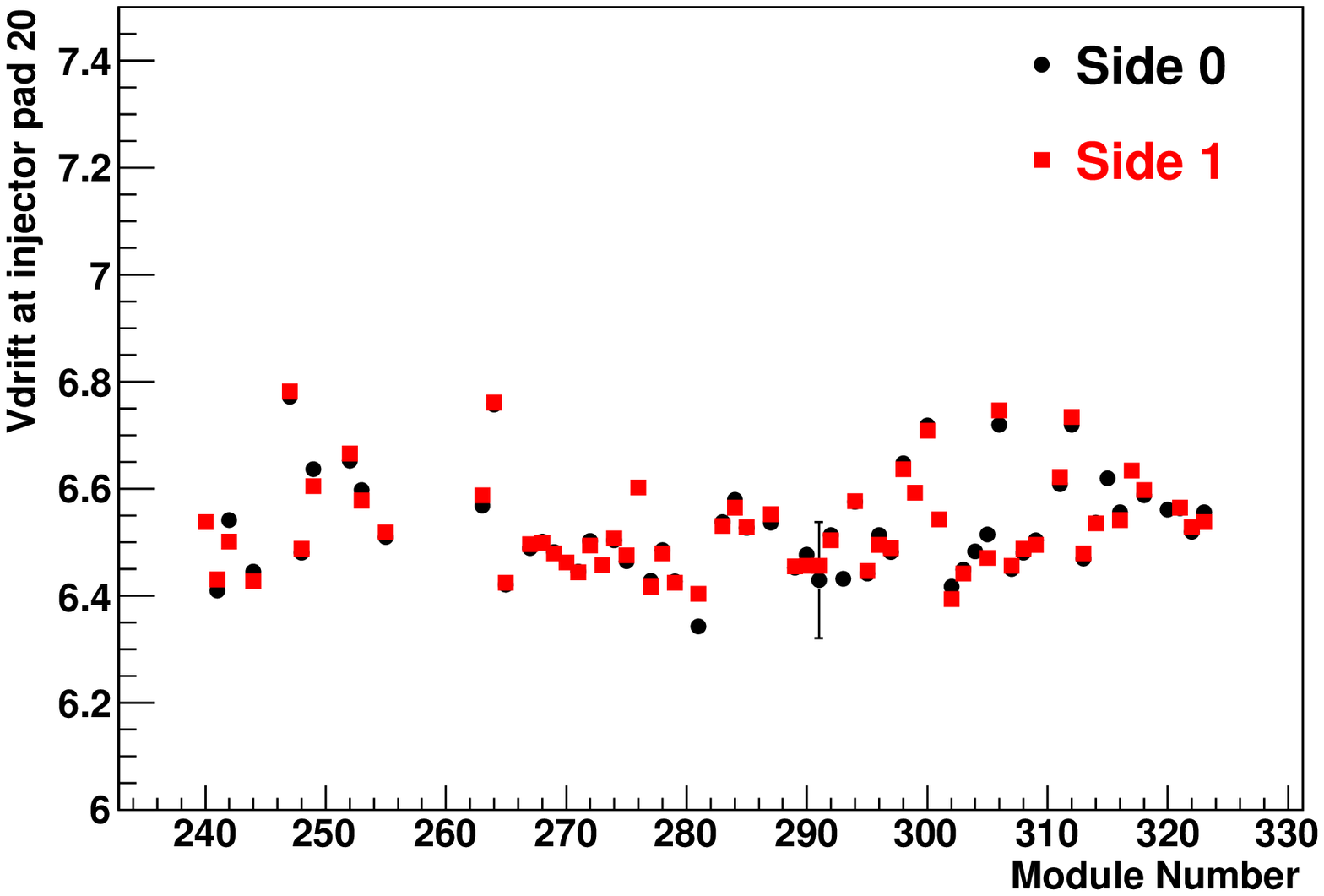}}
&
\mbox{\includegraphics[width=0.48\linewidth]{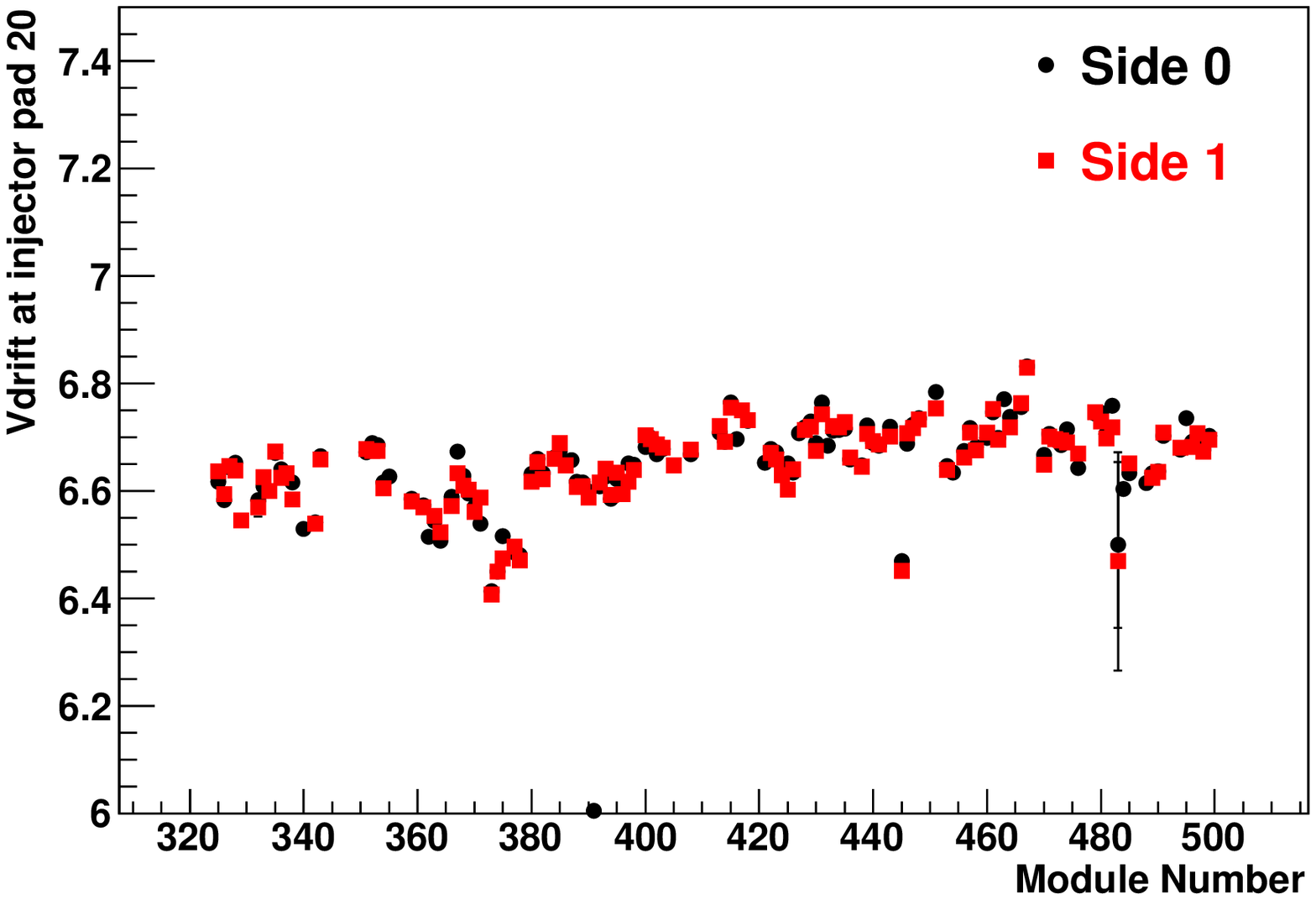}}
\end{tabular}
\end{center}
\caption{Mean drift speed, measured with the same injector pad for the two
sides of a module, as a function of the module number for the Layer 3 (left)
and Layer 4 (right) (where not shown, the error bars are smaller than the
marker size; for two bad anodes the errors are large).}
\label{temp}
\end{figure}

The drift speed depends on the local temperature as $T^{-2.4}$\/. The fact that
heat sources (the voltage dividers) are located at the sensor's edges explains
the anode-by-anode dependence of the drift speed, lower near the warmer edges
and higher in the colder mid region. For the same reason layer 3 has a drift
speed systematically lower than layer 4 because being more internal it has a
slightly higher temperature; this can be seen in Figure \ref{temp}, where the
drift speed for all modules of each layer (measured with the same injector pad)
is shown. Using a nominal value of 1350~$cm^2 V^{-1} s^{-1}$ for the electron
mobility at 293~K in $n$-type silicon with a 3~k$\Omega$~cm resistivity, and
assuming that the mobility scales as $T^{-2.4}$, one can estimate an operating
temperature of about 298~K for Layer 3 and 295~K for Layer 4.

\begin{figure}[t]
\begin{center}
\begin{tabular}{c c}
\mbox{\includegraphics[width=0.47\linewidth]{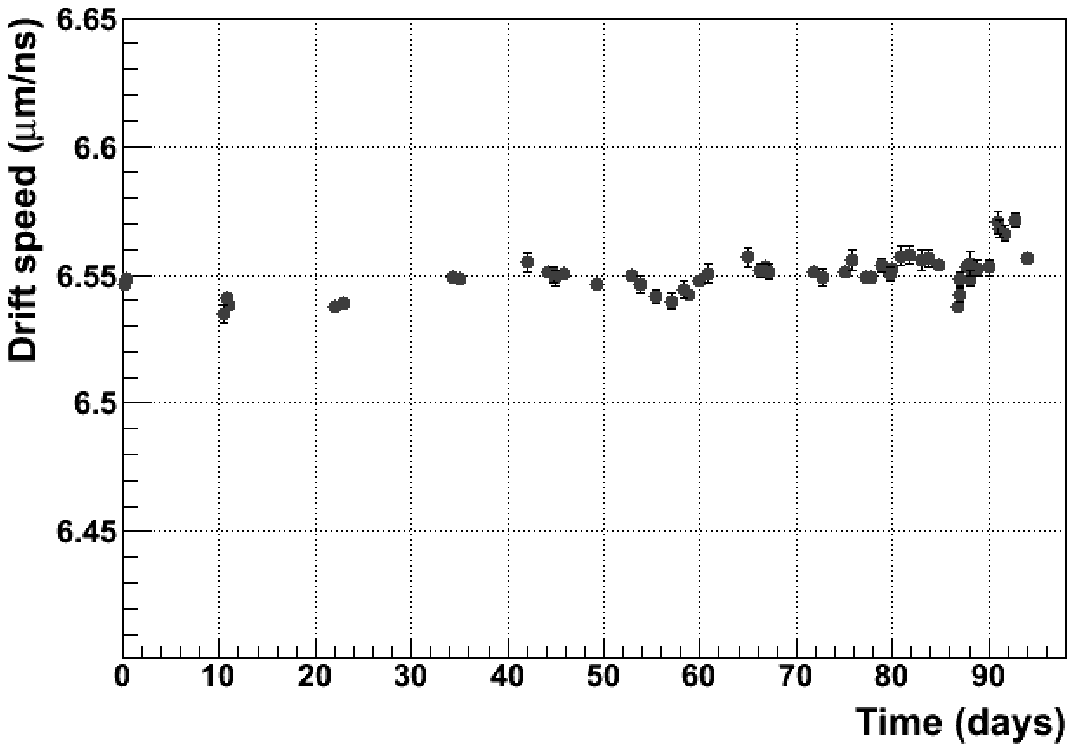}}
&
\mbox{\includegraphics[width=0.47\linewidth]{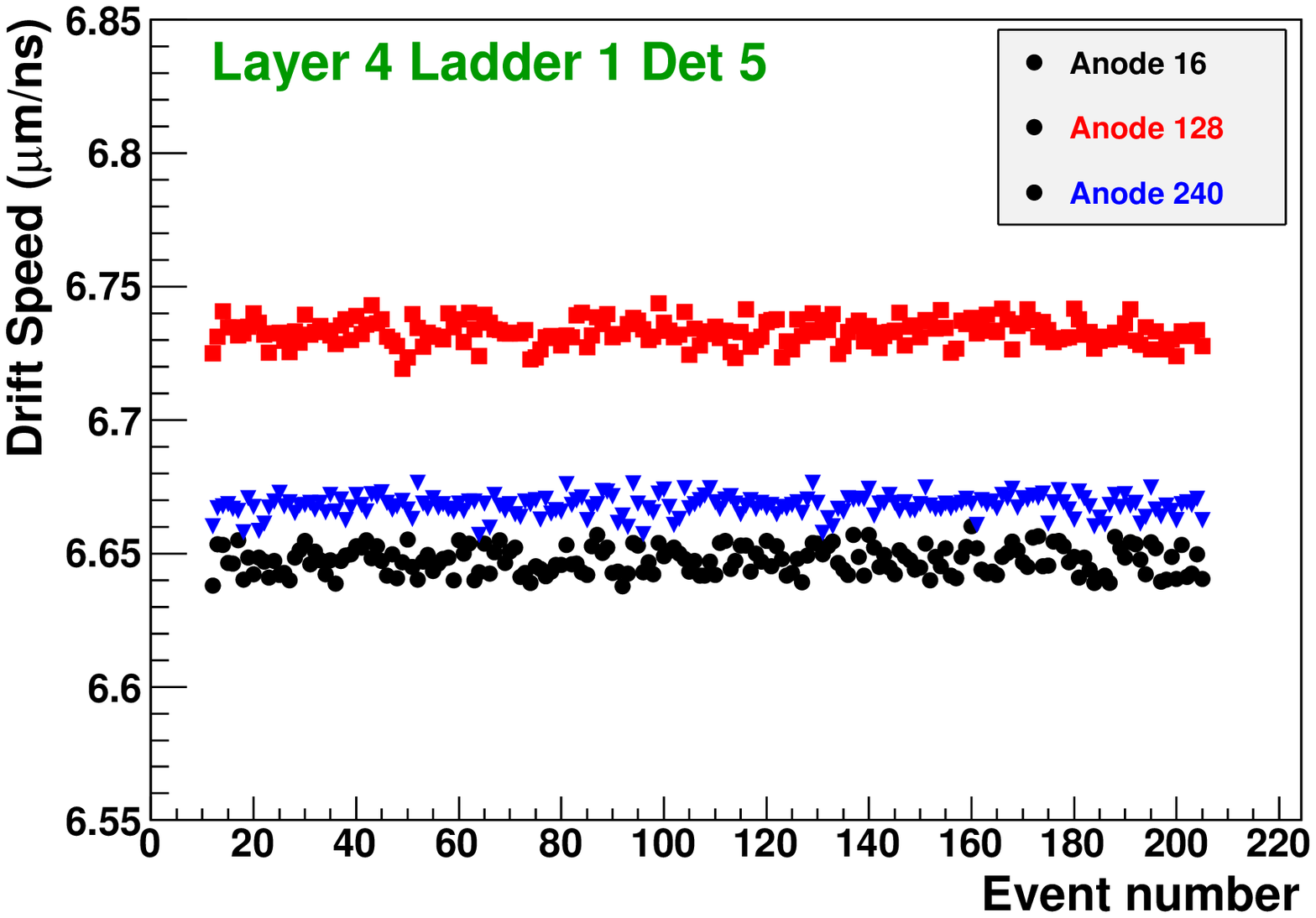}}
\end{tabular}
\end{center}
\caption{Left: Drift speed of one anode as a function of time (days since the
beginning of the 2008 cosmic run). Right: drift speed of three anodes of a half
module as a function of the event number in a special 1 hour run.}
\label{driftime}
\end{figure}

Also the mean drift speed was very stable over the entire period. In the left
half of Figure \ref{driftime} the drift speed of one particular anode is
plotted as a function of the day since the beginning of the cosmic run. In
addition special 1 hour runs were collected in which the average drift speed
was measured every minute; the fluctuations, expressed as the ratio of the
distribution RMS to the mean, were $\simeq$ 0.07\% for most of the modules, as
shown in the right half of Figure \ref{driftime}.

\section{Operation and calibration during the 2008 cosmic run}

\subsection{SOP delay calibration}

As explained in Subs.\ \ref{subsecFERO} a proper delay must elapse between the
trigger arrival time and the memories stop, so that the memory columns which
contain the actual data can send them to ADC. A memory depth of 256 cells with
a 40 MHz sampling rate defines a 6.4 $\mu$s time window, which must contain the
signal of all particles crossing the detector, taking into account that the
maximum drift time is $l_{det}/v_{drift} \simeq 3.5~{\rm cm}/6.5~\mu{\rm m/ns}
\simeq 5.4~\mu$s. This Start Of Process (SOP) delay has to be tuned according
to the trigger type, and during the 2008 data taking this was done during the
T12 injection test. The aim is to center the SDD acquisition time window on the
observed particle times.

\begin{figure}[t]
\begin{center}
\includegraphics[scale=0.48]{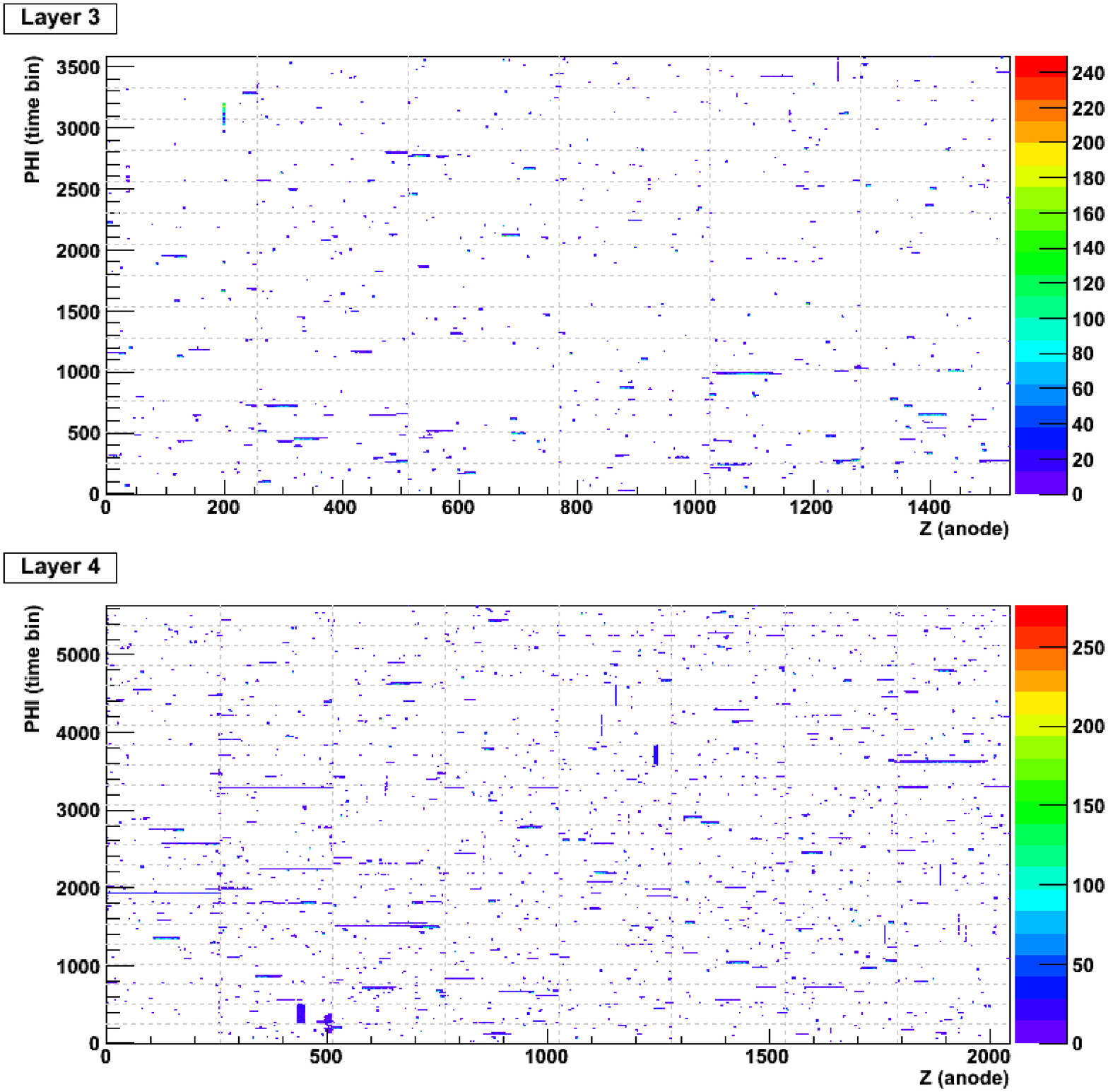}
\end{center}
\caption{Display of an event collected by the SDD during the T12 injection
test. The Z coordinate (along the beam direction) is on the horizontal axis
expressed as anode number, while the vertical axis is the $\varphi$ coordinate,
expressed as time bin (25 ns each). Each blue point is a particle hit;
continuous lines correspond to particles crossing a whole module along the Z
axis.}
\label{eventinject}
\end{figure}

The T12 injection test was one of the LHC commissioning tests. It consisted in
injecting proton bunches into the T12 transfer line, which is the beam line
connecting the SPS to the LHC sector 1--2. The beam was then dumped in the TED
absorber about 300~m upstream of the ALICE detector. Each bunch consisted of
$5\cdot 10^9~p$ at the SPS energy (450 GeV), and was injected every $\sim 50$
s. When the beam was dumped, a shower of secondary particles (about 10
muons/cm$^2$) was produced and a certain fraction reached ALICE illuminating
the central detectors. An example of one such event collected by the SDD is
shown in Figure \ref{eventinject}.

The SOP delay scan was performed for both 40 and 20 MHz sampling rate. The
technique consisted in collecting a certain amount of events with a given SOP
delay value. A preliminary check on the detector occupancy was done, and it
turned out that $\sim 20$ events per SOP delay value were enough. Then the
drift time distribution of the saved data was determined offline using custom
software tools without any event reconstruction. The overall procedure was
repeated in steps of 10 SOP delay units (25 ns) from 110 to 140 (2750 to 3500
ns). The optimal value resulted in the range 120-125 SOP delay units (3000-3125
ns) for both 20 and 40 MHz sample rate.

\begin{figure}[t]
\begin{center}
\includegraphics[scale=0.71]{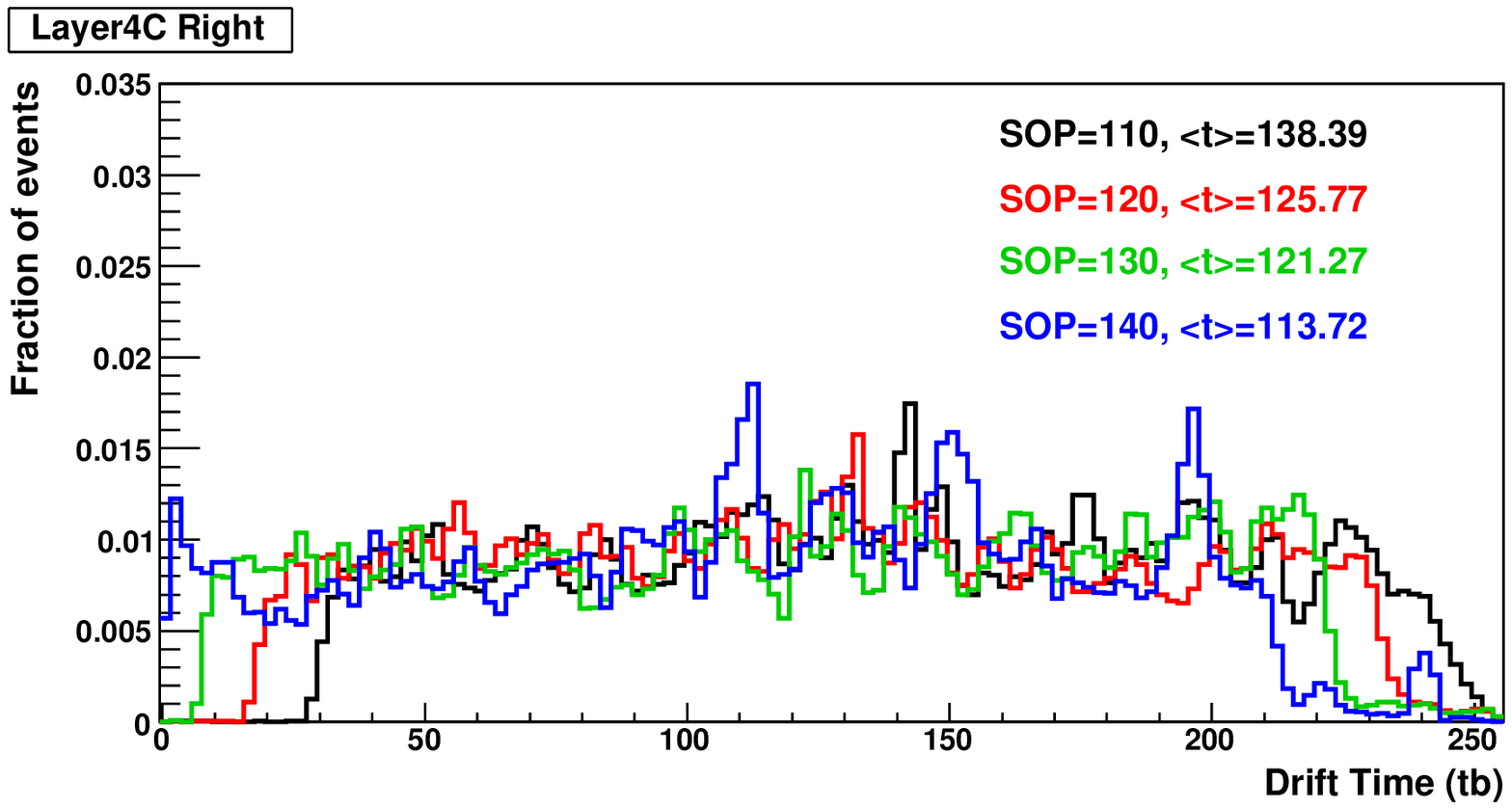}
\end{center}
\caption{Distribution of the particle drift times on Layer 4 (one half-ladder)
sampled at 40 MHz for the four different SOP delay values scanned (peaks are
due to particles crossing modules side by side, as seen in Fig.\ 12).}
\label{sopdelay}
\end{figure}

As an example Figure \ref{sopdelay} shows the distribution of the particle
drift times on Layer 4 sampled at 40 MHz for the four different SOP delay
values examined. Clearly 140 delay units are too many, since the distribution
is truncated at small drift times, while 110 are too few, since the same
distribution is too close to the end of the time bin range. With 120 and 130
SOP delay units the drift time distributions are better centered within the 256
time bins, so 125 units were chosen.

\subsection{Time-zero calibration}

Another parameter measured during the 2008 commissioning is the time-zero, that
is the time offset that has to be subtracted (half-module by half-module) from
all measured drift times to obtain the actual particle time. Two strategies
have been developed based on simulated data.

The first method consists in measuring the minimum drift time from the time
distribution of all measured clusters. The distribution of all drift times is
constructed. Then the rising part of the distribution at small drift times,
which in the ideal case would be a sharp rise, is fitted with an error
function. The time zero is then extracted from the fit. This procedure provides
a direct measurement of the time zero and does not require any other
calibration parameter such as the drift speed; moreover it is an SDD standalone
measurement without any use of information from the other ITS subdetectors. But
it has the disadvantage of requiring a large statistics of reconstructed
points.

The second method consists in measuring the track-to-point residuals. Muon
tracks are fitted in SPD and SSD, using at least 5 points, and the residuals
between the track crossing point taken as reference and the cluster coordinate
in SDD are computed. The time offset is extracted by exploiting the opposite
sign of the residuals in the two detector sides (see Fig.\
\ref{fig:sddlayout}): an uncalibrated time offset would lead to an
over/underestimation of the drift path, and therefore to residuals of opposite
signs in the two sides. The distance between the two peaks of the distribution
of residuals in the two detector sides corresponds to $2 v_{drift} t_0$\/, thus
leading to the determination of the time zero \cite{alignpaper,prinord09}. This
procedure requires less statistics, but relies on calibration parameters (the
drift speed and the correction maps, see Sec. 2); moreover, being based on
track reconstruction in ITS, it might be biased by SPD and/or SSD
misalignments.

As a first approximation the time zero can be considered equal for all modules
apart from corrections for different cable lengths: actually a significant
difference ($\sim 25$ ns) between the modules in the half ladders located on
positive and negative $z$ has been found due to the different optical fiber
length (about 6 m).
 
\subsection{Other calibrations}

\begin{figure}[t]
\begin{center}
\includegraphics[scale=0.75]{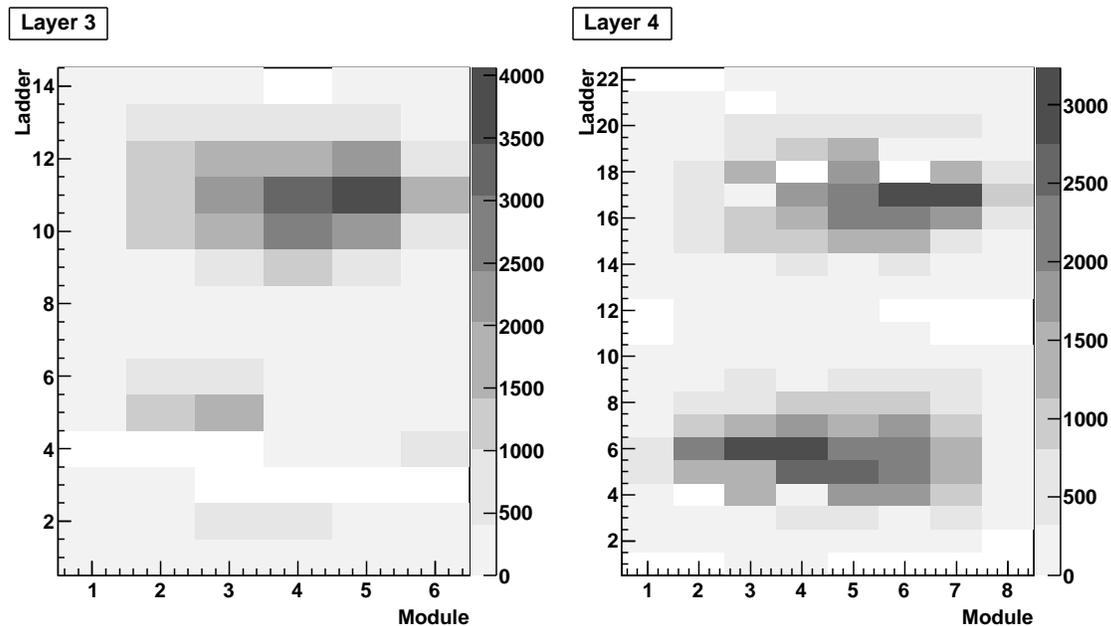}
\end{center}
\caption{Detector occupancy with reconstructed cosmic muons for Layer 3 (left)
and Layer 4 (right). Layer 3 shows less statistics because it was off for about
one month due to hardware problems to a read-out electronic crate.}
\label{align}
\end{figure}

During the commissioning with cosmic ray data other important calibrations were
performed, namely the charge calibration and the detector alignment. From the
charge distribution of the cosmic muons it is possible to extract the
conversion factor from ADC counts to keV, using the most probable value of
energy deposition in the detector thickness. Reconstructing the muon tracks
(and successively also the particle tracks from the beam-gas and the first p-p
interactions) it is possible to align with each other the ITS subdetectors and
subsequently the ITS and the TPC. These further calibrations are described
elsewhere in other papers \cite{alignpaper,prinord09}. In Figure
\ref{align} the module-ladder map of reconstructed cosmic muons for SDD Layer 3
and Layer 4 is shown; clearly the horizontal modules (around ladder 4 and 11
for Layer 3 and ladder 6 and 17 for Layer 4) have much more entries that the
vertical or quasi-vertical ones.

\section{Conclusions}

During the 2008 commissioning run with cosmic muons the Silicon Drift Detectors
of the ALICE experiment were regularly calibrated in order to determine the
baselines, the noise, the gain and the drift speed. All monitored parameters
proved to be stable during the entire period (within 1\% for the drift speed).
Supplementary calibrations for the SOP delay and the time offset were also
performed by exploiting the injection test and the reconstructed muon tracks.
The detector was fully operational and ready for the planned p-p data taking.

After the major experiment shutdown during the winter, the detector was brought
up for the new 2009 commissioning run. Similar calibrations are being performed
to monitor the detector performance and measure its working parameters.

\acknowledgments
We wish to thank the people helping us during the construction phase at the
INFN Technological Laboratory in Turin.
For the mechanical part: G.\ Alfarone, F.\ Borotto, F.\ Cotorobai, R.\ Panero,
L.~Simonetti.
For the electronics: P.\ Barberis, M.\ Mignone, F.\ Rotondo.
We wish to thank also the people involved in the assembly and bonding of the
SDD modules, whose work was fundamental for the SDD project: F.\ Dumitrache,
B.\ Pini, O.\ Chykalov, L.\ Klimova, L.\ Ruzhystka, I.\ Tymchuk and the team
from SRTIIE, Kharkov, Ukraine. We warmly thank also M.\ Bondila, G.\ Casse,
D.~Cavagnino, W.\ Dabrowski, G.\ Herrera, E.\ Lopes Torres, S.\ Martoiu, R.\
H.\ Montoya, V.\ Pospisil, A.\ Zampieri, and all students who graduated working
on the SDD project.

This work was partly supported by the Ministry of Education of the Czech
Republic under Grants N.\ LA08015 and LA07048.

This work was partly supported by the U.\ S.\ National Science Foundation under
Grant N.\ PHY-0653432308.

This work was partly supported by the European Union, the Regione Autonoma
Valle d'Aosta and the Italian Ministero del Lavoro e della Previdenza Sociale.

\end{document}